\documentclass[11pt]{article}
\usepackage[a4paper,margin=0.90in]{geometry}
\usepackage[T1]{fontenc}
\usepackage[utf8]{inputenc}
\usepackage{lmodern}
\usepackage{amsmath,amssymb, bm, url, amsfonts, xcolor, graphicx, authblk, tcolorbox, breqn}
\usepackage{booktabs, fontawesome5}
\usepackage{enumitem}
\usepackage{tikz}
\usetikzlibrary{arrows.meta, positioning, fit, backgrounds, calc}
\usepackage[colorlinks=true,linkcolor=blue,urlcolor=blue,citecolor=blue]{hyperref}
\usepackage{tocloft}
\tcbuselibrary{skins, breakable}

\newcommand{\greendiamond}{{\color{green}\blacklozenge}}
\newcommand{\magentadiamond}{{\color{magenta}\blacklozenge}}

\newtcolorbox{terminal}{
  colback=gray!8,
  colframe=gray!70!black,
  fonttitle=\bfseries\small,
  title=terminal:$\sim$\$,
  breakable,
  left=6pt, right=6pt
}

\newtcolorbox{macOS}{
  colback=gray!8,
  colframe=gray!70!black,
  fonttitle=\bfseries\small,
  title=macOS,
  breakable,
  left=6pt, right=6pt
}

\newtcolorbox{Windows}{
  colback=gray!8,
  colframe=gray!70!black,
  fonttitle=\bfseries\small,
  title=Windows,
  breakable,
  left=6pt, right=6pt
}

\newtcolorbox{user}{
  colback=magenta!8,
  colframe=magenta!70!black,
  fonttitle=\bfseries\small,
  title=User>,
  breakable,
  left=6pt, right=6pt
}

\newtcolorbox{agent}{
  colback=cyan!8,
  colframe=cyan!50!black,
  fonttitle=\bfseries\small,
  title=BSM\_Agent>,
  breakable,
  left=6pt, right=6pt
}

\newtcolorbox{pdf}{
  colback=white,
  colframe=yellow!50!black,
  fonttitle=\bfseries\small,
  title=Generated PDF>,
  breakable,
  left=6pt, right=6pt
}

\definecolor{darkergreen}{HTML}{006400} 

\title{
    \bf Large Language Model-Assisted Framework \\for BSM Model Building \\[0.5em] 
    \mbox{\rule{\linewidth}{1pt}}
}
\author{
    \bf{\Large{Shaikh Saad}} \\ \vspace{0.1em}
     {\small Jožef Stefan Institute, Jamova 39, P.\ O.\ Box 3000, SI-1001 Ljubljana, Slovenia} \\ 
    \texttt{\color{magenta}shaikh.saad@ijs.si} \\ \vspace{0.5em}
    \mbox{\rule{\linewidth}{1pt}}
}
\date{}

\begin{document}
\maketitle

\begin{abstract}
Recent advances in artificial intelligence (AI), particularly large language models (LLMs), have created new opportunities for natural-language interaction with scientific software, but reliable theoretical model building still requires deterministic symbolic calculations. We present \texttt{bsm_agent}, an open-source symbolic framework for beyond the Standard Model (BSM) model building that combines a deterministic physics backend with an LLM chat interface. Starting from the SM field content and a user-specified set of additional scalars and/or fermions, the package constructs renormalizable Lagrangian, performs gauge-anomaly checks, expands operators into component fields, and derives electroweak symmetry breaking stationary conditions and tree-level mass matrices. The key novelty of the framework is that all of these tasks are performed automatically once the user specifies the quantum numbers of the new fields through a natural-language interface, eliminating the need for manual model construction. The symbolic calculations are performed entirely by the Python backend to ensure the correctness and reproducibility of the physics results; the LLM is used only as an orchestration layer that interprets natural-language requests, manages confirmation steps for ambiguous inputs, triggers backend tools, and formats report-ready summaries. The package supports three provider classes: local Ollama inference, remote self-hosted model servers accessed through the implemented remote provider interface, and commercial hosted APIs via OpenAI and Anthropic. This separation between conversational control and deterministic computation preserves reproducibility while making interactive BSM model construction substantially more convenient.  \href{https://github.com/saad-hep/bsm_agent}{\faGithub~bsm_agent}
\end{abstract}

\clearpage
\tableofcontents
\clearpage
\section{Introduction}
A central goal of particle physics is to construct theoretical models that describe and explain the fundamental constituents of nature and their interactions. At present, the Standard Model (SM)~\cite{Glashow:1961tr,Weinberg:1967tq,Salam:1968rm} of particle physics provides the most successful description of elementary particles and their interactions. The SM is based on a gauge-invariant Lagrangian containing a large number of interaction terms and has been remarkably successful in explaining a wide range of experimental observations.

Despite its tremendous success, the SM is known to be incomplete. Among its most important shortcomings are: (i) the prediction of massless neutrinos, whereas neutrino-oscillation experiments have firmly established that neutrinos possess non-zero masses; (ii) the inability to account for the observed matter--antimatter asymmetry of the Universe; and (iii) the absence of a viable dark matter candidate. These observations strongly suggest the existence of physics beyond the SM (BSM).

One of the simplest approaches to BSM model building is to extend the particle content of the SM through the addition of new scalar and/or fermionic fields while leaving the gauge structure unchanged. This class of models encompasses a large variety of phenomenologically interesting scenarios and serves as a common starting point for theoretical investigations.

Constructing such models, however, can be a laborious and error-prone process. For a given set of new fields, one must identify all renormalizable gauge-invariant operators, verify gauge-anomaly consistency, derive the full Lagrangian, expand operators into component fields for phenomenological studies, and obtain the corresponding mass matrices after electroweak symmetry breaking. As the number of new fields increases, the number of allowed interaction terms grows rapidly, making it increasingly difficult to ensure that all relevant operators are included in the Lagrangian and that only linearly independent operators are retained.

To address these challenges, we introduce \texttt{bsm_agent}, an open-source Python framework for symbolic BSM model building. The package combines a deterministic physics backend with a large-language-model~\cite{zhao2026survey} (LLM) chat interface, enabling users to construct and analyze models through natural-language interactions while retaining the reliability and reproducibility of symbolic computations.  Given the SM field content and a user-specified set of additional scalar and/or fermionic fields, the framework automatically generates the complete renormalizable operators, performs gauge-anomaly diagnostics, expands the Lagrangian into component fields, inserts vacuum expectation values (VEVs) of color and charge neutral states, lists the stationary conditions, and derives tree-level mass matrices,  requiring no manual model construction beyond the specification of the new fields' quantum numbers through a natural-language chat interface. All physics calculations are carried out by the symbolic backend, while the LLM serves exclusively as an orchestration and user-interface layer.

The primary objective of this software package is to enhance the accessibility of intricate model-building methodologies, rendering advanced theoretical frameworks manageable for students and pedagogical use while maintaining the rigor required by research professionals. To provide a fast-track for experienced model builders eager to start implementation, the foundational and interface-level chapters (Sections~\ref{sec:02}-\ref{sec:04}) can be omitted, with Section~\ref{sec:05} serving as the practical entry point. Furthermore, Appendix~\ref{summarization} provides the quickest way for readers to begin using the package immediately.

This paper is organized in the following way. Section~\ref{sec:02} outlines the overarching architecture and fundamental design principles governing our framework. Section~\ref{sec:03} deep dives into the deterministic symbolic backend, covering field content extensions, gauge-anomaly validation checks, automated Lagrangian generation, and the mechanics of component field expansion alongside electroweak symmetry breaking. Section~\ref{sec:04} introduces the LLM-assisted interaction layer, detailing its tool-calling architecture, session state workflows, and the supported provider classes for local, remote self-hosted, and commercial API backends. Section~\ref{sec:05} walks through the core Standard Model implementations, demonstrating how to generate comprehensive reports, expand the Lagrangian, and derive mass matrices. Section~\ref{sec:06} provides a practical, step-by-step walkthrough for building a new custom physics model from scratch. Section~\ref{sec:07} catalogues the comprehensive features of the package, specifying fields, scalars, quantum number assignments, Yukawa configurations, and normalization conventions. Section~\ref{sec:08} presents a complete case study centered on a Leptoquark model, demonstrating end-to-end model construction, PDF generation, full Lagrangian expansion across sectors, and mass matrix extraction. Section~\ref{sec:09} summarizes our findings and offers concluding remarks on future developments.

\begin{figure}[th!]
\centering
\includegraphics[width=1\textwidth]{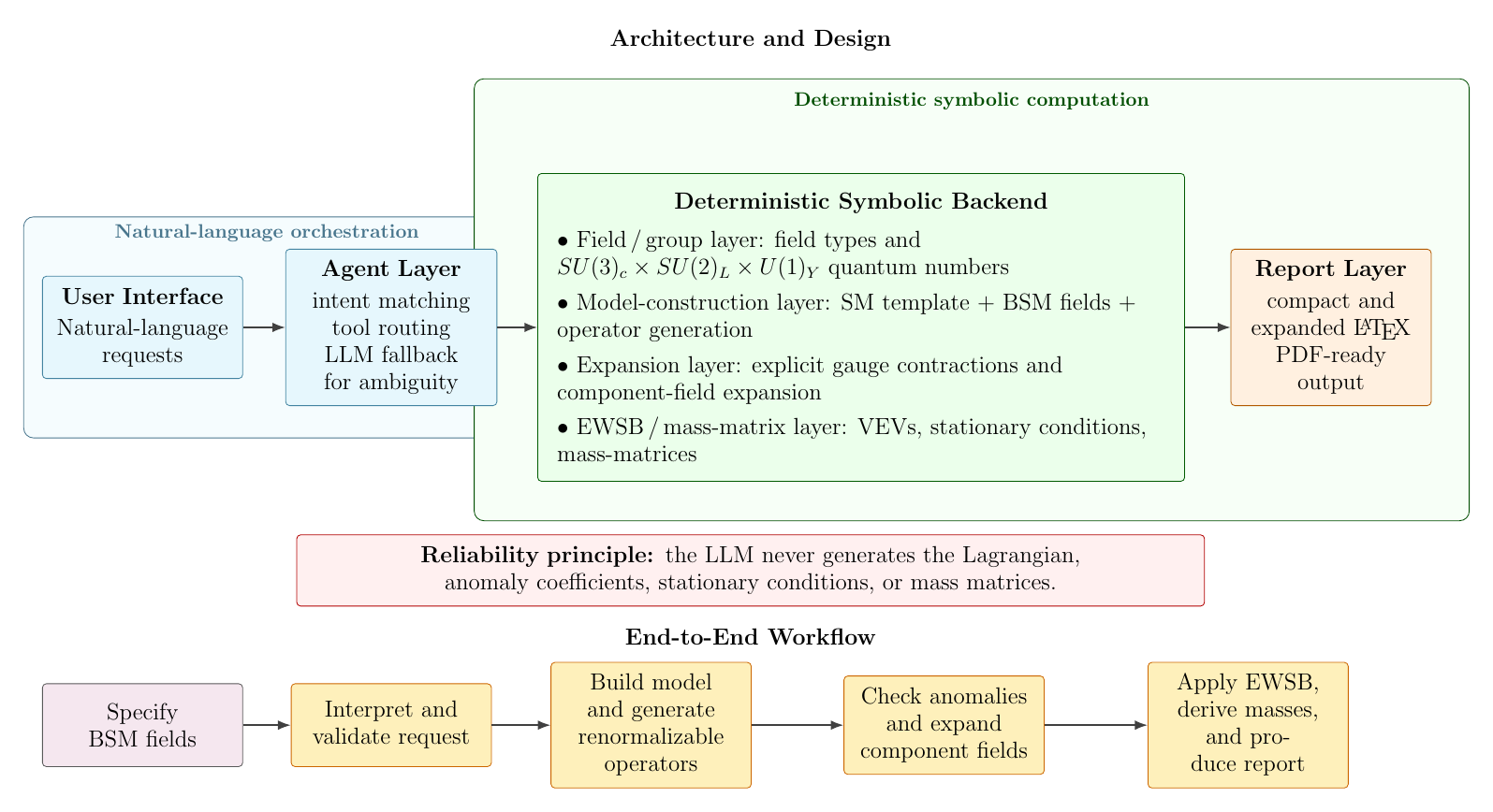}
\caption{Architecture and workflow of \texttt{bsm\_agent}. Natural-language user
requests are handled by the agent layer, which performs intent matching, tool
routing, and LLM-assisted interpretation only when needed. All physics outputs
are produced by the deterministic symbolic backend: field and representation
handling, renormalizable operator generation, component-field expansion,
electroweak-symmetry-breaking analysis, mass-matrix derivation, and report-ready
\LaTeX{} generation.}
\label{fig:architecture}
\end{figure}

\section{Architecture and Design Principles}\label{sec:02}

LLMs and AI agents are rapidly becoming integral components of modern scientific research. Across disciplines, they are being used for literature discovery, scientific synthesis, code generation, hypothesis formulation, workflow automation, and data analysis. Recent large-scale bibliometric studies have shown that AI adoption has expanded across virtually all scientific disciplines and is now a pervasive component of contemporary research practice \cite{hajkowicz2023artificial}. More broadly, the emerging field of AI for Science  demonstrates how machine learning, foundation models, and LLMs can accelerate discovery across quantum, atomistic, materials, biological, and physical systems \cite{zhang2025artificial}. Several surveys have highlighted the growing role of LLMs in scientific reasoning, synthesis, explanation, and interdisciplinary research workflows \cite{boyko2023interdisciplinary,zheng2023large}. Beyond passive assistance, recent work on LLM-based agents argues that AI systems can autonomously plan, execute, and evaluate complex scientific workflows while maintaining human oversight \cite{wang2024survey,Li:2026krn}. Such capabilities are increasingly being explored in scientific domains ranging from climate science and Earth-system modeling to biomedical and physical sciences \cite{millison2026state}. In high-energy physics, agentic AI systems have already demonstrated the ability to support simulation workflows, automate event generation, reproduce published analyses, and assist both novice and expert researchers \cite{plehn2026madagents,moreno2026autonomousphysics,colliderbench2026,Esmail:2026jpb,Qiu:2026iby,Agrawal:2026lvg,Menzo:2026qrl,Faroughy:2026dkj,Desai:2026nmx,Wang:2026jjn}. These developments suggest that conversational AI interfaces are evolving from simple question-answering tools into scientific collaborators capable of reducing technical barriers, improving reproducibility, and accelerating the pace of discovery.  The package presented in this work demonstrates how an LLM can serve as a natural-language interface to established particle-physics software, enabling access to complex calculations with minimal technical overhead while leaving the underlying computations to dedicated physics frameworks.

The package is organized into a small number of modules that map naturally onto the BSM model-building workflow; see Fig.~\ref{fig:architecture}.

\begin{enumerate}[leftmargin=*]

\item \textbf{Field and group layer.}
Field classes and group-representation utilities define Standard Model quantum numbers, distinguish scalar and Weyl-fermion field types, and provide the corresponding representation data under the Standard Model gauge group $SU(3)_c \times SU(2)_L \times U(1)_Y$.

\item \textbf{Model-construction layer.}
A model container initializes the Standard Model field content and can be extended by  additional scalar and fermionic fields. This layer is responsible for generating the complete renormalizable gauge-invariant operator basis. The current implementation focuses on non-supersymmetric extensions of the Standard Model.

\item \textbf{Expansion layer.}
Gauge-invariant operators are expanded into explicit component fields using algorithms that perform the required $SU(2)$ and $SU(3)$ index contractions.

\item \textbf{EWSB and mass-matrix layer.}
Neutral scalar fields are assigned VEVs. The package then derives the corresponding stationary conditions and constructs the tree-level mass matrices by collecting the relevant terms of the shifted Lagrangian.

\item \textbf{Report layer.}
Compact and fully expanded \LaTeX{} representations of the model are generated directly from the symbolic backend.

\item \textbf{Agent layer.}
The agent interface connects the model-building and report-generation tools to a conversational backend and manages the interaction workflow. Users communicate with the framework through natural-language requests. Whenever possible, requests are resolved through deterministic command matching; ambiguous or complex queries may be delegated to an LLM for interpretation before invoking the symbolic backend.
\end{enumerate}

This separation of responsibilities is central to the reliability of the framework. The LLM never generates the Lagrangian, anomaly coefficients, stationary conditions, or mass matrices. Instead, it serves exclusively as an orchestration layer that invokes deterministic symbolic computations performed by the backend.

\section{Deterministic Symbolic Backend} \label{sec:03}
\subsection{Field Content and Model Extension}
The symbolic workflow begins with a Standard Model template and a list of extra fields. Each field is specified by its name, whether it is a scalar or a Weyl fermion, its $SU(3)$ representation, its $SU(2)$ representation, its hypercharge, and whether the field is real. The model container prevents duplicate field names and combines the SM content with the new BSM degrees of freedom into a single internal representation. The automatic operator generator is restricted to operators of mass dimension less than or equal to four, i.e., renormalizable. This is intentional: it keeps the generated operator basis finite, model independent, and directly useful for tree-level studies of the scalar and fermion sectors. 

The current implementation is within the SM gauge group,
$SU(3)_c\times SU(2)_L\times U(1)_Y$,
and supports the following classes of gauge representations $(Z,X,Y)$ under the SM:
\begin{itemize}

\item[$\greendiamond$] Arbitrary values of the $U(1)_Y$ hypercharge, $Y$.

\item[$\greendiamond$] $SU(2)_L$ representations of dimensions $X=1$, $2$, $3$, $4$, $5$, $6$, and $7$.

---This choice is motivated by minimal dark matter scenarios~\cite{Cirelli:2005uq}, in which the lowest-dimensional viable candidates correspond to a fermionic quintuplet and a scalar septuplet. These minimal realizations do not require the introduction of any additional symmetries beyond those of the SM to ensure dark matter stability.

---While the framework supports $SU(2)_L$ representations up to dimension $7$, the computational complexity of scalar-potential expansion increases rapidly for models containing multiple higher-dimensional scalar multiplets. The current implementation therefore focuses on the phenomenologically most common scenarios, for which complete symbolic results can be generated within practical runtimes. In particular, for scalar multiplets with $SU(2)_L$ dimension $X>3$, the package supports models containing a single field with quantum numbers $(1,X,Y)$, including complete expansion of the Lagrangian into component fields.

\item[$\greendiamond$] $SU(3)_c$ representations  $Z=1$, $3$, $\overline{3}$, $6$, $\overline{6}$, and $8$.

---This choice is motivated by realistic $SO(10)$~\cite{Fritzsch:1974nn} grand-unified 
constructions, where commonly used representations such as $10$, $16$, $45$, $54$, $120$, $126$, and $210$ do not contain color representations larger than the adjoint octet under $SU(3)_c$.

---Due to the computational complexity associated with the symbolic expansion of the Lagrangian, the current implementation supports fields transforming as $(Z,X,Y)$ with $X\leq 3$.

\end{itemize}

Taken together, this broad range of gauge representations allows users to construct a wide variety of phenomenologically interesting extensions of the Standard Model. The current restrictions on higher-dimensional representations arise from computational considerations associated with symbolic expansion and will be progressively relaxed in future versions of the package.

It is important to emphasize that \texttt{bsm_agent} is complementary to existing packages such as \texttt{SARAH}~\cite{Staub:2008uz}, \texttt{FeynRules}~\cite{Christensen:2008py}, and \texttt{Susyno}~\cite{Fonseca:2011sy},  which provide powerful frameworks for implementing and analyzing particle-physics models. In these tools, the user typically specifies the model definition, including the relevant interaction structure or Lagrangian, from which physical quantities such as interaction vertices, mass spectra, and other observables can be derived. By contrast, the primary goal of \texttt{bsm_agent} is to automate an earlier stage of the model-building workflow. Starting only from the Standard Model field content and the quantum numbers of newly introduced scalar and/or fermionic fields, the framework automatically constructs the complete renormalizable gauge-invariant operator basis and derives the corresponding Lagrangian. This significantly reduces the manual effort required during the initial stages of BSM model construction. At present, the framework is restricted to extensions that preserve the Standard Model gauge symmetry and introduce additional scalar and/or fermionic fields. Extensions involving enlarged gauge groups constitute a natural direction for future development but are beyond the scope of the present work.

\subsection{Gauge-Anomaly Checks}
Before discussing operator generation, it is useful to note that anomaly checking is implemented natively in the backend. The code computes the standard anomaly coefficients by summing over the left-handed Weyl fermions present in the model. The reported coefficients are
\begin{align}
\mathcal{A}_{SU(3)^3}, \qquad
\mathcal{A}_{SU(2)^2U(1)}, \qquad
\mathcal{A}_{SU(3)^2U(1)}, \qquad
\mathcal{A}_{\text{grav}^2U(1)}, \qquad
\mathcal{A}_{U(1)^3}.
\end{align}
For each fermion representation, the appropriate Dynkin-index and hypercharge weights are accumulated, and the final report states whether all five coefficients vanish. 

If non-vanishing anomaly coefficients are detected, the package may suggest the addition of conjugate fermionic representations associated with the fields responsible for the anomaly. Such suggestions are intended only as a convenience for exploratory model building and should not be interpreted as a complete solution to the anomaly-cancellation problem. Indeed, gauge anomalies can be cancelled in many different ways, and the introduction of conjugate representations represents only one possible strategy.

\subsection{Automatic Generation of the Renormalizable Lagrangian}
Once the field content is fixed, the package constructs the full renormalizable gauge-invariant operator basis. Importantly, the user is not required to specify any interaction terms manually. The complete renormalizable Lagrangian is derived directly from the field content and gauge quantum numbers. The generation step is centered on the model container and proceeds by operator class.

\paragraph{Scalar sector.}
All scalar monomials containing up to four scalar fields are enumerated using combinations with replacement. For each candidate monomial, the backend checks whether the product contains at least one gauge singlet and whether its total mass dimension is at most four. Allowed singlets are then added to the scalar potential. The corresponding coefficient symbols are assigned automatically with prefixes that reflect the degree of the operator, namely
\begin{align}
\xi^3, \qquad m^2, \qquad \mu, \qquad \lambda,
\end{align}
for linear, quadratic, cubic, and quartic terms, respectively. If a given field product admits several independent gauge contractions, the package indexes them explicitly. For example, consider a scalar quartic interaction with coupling $\lambda_i$. If three independent gauge-invariant contractions exist for this interaction, the package labels the corresponding couplings as $\lambda_{i,c1}$, $\lambda_{i,c2}$, and $\lambda_{i,c3}$.

For every scalar field added to the model, the package automatically assigns a symbolic identifier ${\color{blue}\phi_i}$ (for clarity, the package uses \textcolor{blue}{BLUE} to denote any BSM scalar field), with the index $i$ distinguishing different scalar species.

\paragraph{Fermion mass sector.}
All bilinears constructed from Weyl fermions are tested for gauge invariance. Whenever a singlet exists, a fermion-mass operator is generated with an automatically named coefficient $M_i$ or $M_{i,c}$ when several contractions are present.

For every Weyl fermion added to the model, the package automatically assigns a symbolic identifier ${\color{red}\psi_i}$ (for clarity, the package uses \textcolor{red}{RED} to denote any BSM fermion field), with the index $i$ distinguishing different Weyl-fermion species.

\paragraph{Yukawa sector.}
The Yukawa sector is generated by combining two Weyl fermions with one scalar factor and again testing for singlet multiplicity. If the product is gauge invariant, the corresponding Yukawa operator is added with a coefficient $Y_i$ or $Y_{i,c}$.

It is important to note that, in all Yukawa terms, the family/generation indices are suppressed.

\paragraph{Hermitian-conjugate handling and deduplication.}
The generator explicitly avoids double counting. Operators are ordered canonically, compared against their Hermitian conjugates, and only added once when the conjugate term is not independent. The resulting Lagrangian therefore has a deterministic and stable symbolic representation.

The output of this stage is a symbolic Lagrangian object partitioned into scalar-potential, Yukawa, fermion-mass, and gauge-kinetic sectors. This representation serves as the common input for component-field expansion, electroweak-symmetry-breaking analysis, mass-matrix derivation, and report generation.

\subsection{Computational Scaling with Large Gauge Representations}
The runtime of the automatic model-construction stage depends not only on the number of new fields but also on the sizes of their gauge representations. In particular, fields transforming in higher-dimensional $SU(2)_L$ and/or $SU(3)_c$ representations generally require longer execution times than singlets, doublets, or triplets. This behavior is expected, because the operator generator must test a larger set of candidate renormalizable field combinations and, for each candidate, determine whether one or more independent gauge-singlet contractions exist.

The underlying group-theoretic cost also increases rapidly with representation size. Larger multiplets produce larger tensor-product spaces, a greater number of admissible index contractions, and, in some cases, a higher multiplicity of independent invariant tensors. The scalar sector is particularly demanding in this respect, since repeated scalar fields require an additional projection onto the Bose-symmetric subspace in order to retain only linearly independent operators. Consequently, even models containing only a small number of new fields can exhibit noticeably longer runtimes when those fields belong to large non-Abelian representations. This is therefore a computational consequence of the exact symbolic construction of the operator basis, rather than an indication of any inconsistency in the model itself.

To mitigate this growth, the implementation includes several representation-aware optimizations for the cases that are known to arise frequently in phenomenological studies. In the weak sector, repeated identical scalar multiplets are treated in a reduced Bose-symmetrized basis rather than in the full tensor-product basis, which is particularly beneficial for large real $Y=0$ multiplets such as quintuplets and septuplets. For selected self-interaction structures of real weak triplets, quintuplets, and septuplets, the expansion backend also uses direct closed-form constructions of the corresponding invariant combinations. In the color sector, octet contractions are handled through a dedicated adjoint-basis construction together with pruning of redundant contraction patterns. These optimizations substantially improve performance for several computationally demanding cases, although the runtime can still increase appreciably for models containing large non-Abelian representations. The performance benchmarks for model construction and Lagrangian expansion across various field contents are summarized in Table~\ref{tab:model_benchmarks}, detailing both the computational runtime and the structural complexity of each model.

\begin{table}[htbp]
\centering
\caption{Model Construction and Lagrangian Expansion Runtime Benchmarks}
\label{tab:model_benchmarks}
\small
\renewcommand{\arraystretch}{1.5} 
\begin{tabular}{lccc}
\hline
\textbf{Field Content} & 
\textbf{\begin{tabular}[c]{@{}c@{}}Construction\\ Time (s)\end{tabular}} & 
\textbf{\begin{tabular}[c]{@{}c@{}}Expanding\\ Time (s)\end{tabular}} & 
\textbf{\begin{tabular}[c]{@{}c@{}}Kinetic +\\ Fermion Mass +\\ Scalar Potential +\\ Yukawa = Total\end{tabular}} \\
\hline
$(1,1,0)_F$~\cite{Minkowski:1977sc} & 0.1 & 1.5 & $10 + 1 + 2 + 4 = 17$ \\
\hline
$(1,3,0)_S^R + (1,3,1)_S$~\cite{Georgi:1985nv,Kundu:2021pcg} & 0.1 & 2.0 & $11 + 0 + 16 + 4 = 31$ \\
\hline

$(1,3,1)_F + (1,3,-1)_F + (1,4,3/2)_S$~\cite{Babu:2009aq} & 0.1 & 2.0 & $12 + 1 + 8 + 5 = 26$ \\
\hline

$(3,1,-1/3)_S + (3,1,-4/3)_S + (3,2,7/6)_S +$ & 1.5 & 6.2 & $14 + 0 + 73 + 14 = 101$ \\
$(3,2,1/6)_S + (3,3,-1/3)_S$~\cite{Buchmuller:1986zs,Dorsner:2016wpm,Crivellin:2021ejk} & & & \\
\hline

$(1,7,0)_S^R$~\cite{Cirelli:2005uq} & 0.1 & 1.7 & $10 + 0 + 6 + 3 = 19$ \\
\hline
$(1,7,2)_S$~\cite{Hisano:2013sn,Alvarado:2014jva} & 0.1 & 3.0 & $10 + 0 + 9 + 3 = 22$ \\
\hline

$(6,1,4/3)_S + (6,1,-2/3)_S + (6,1,1/3)_S$~\cite{Ma:1998pi} & 0.5 & 12.2 & $12 + 0 + 27 + 7 = 46$ \\
\hline

$(8,4,1/3)_F + (8,4,-1/3)_F$ & 0.1 & 2.4 & $11 + 1 + 2 + 3 = 17$ \\
\hline

$(8,2,1/2)_S$~\cite{Manohar:2006ga} & 48.3 & 10.5 & $10 + 0 + 14 + 5 = 29$ \\
\hline

$(8,2,1/2)_F + (8,2,-1/2)_F + (8,2,1/2)_S$ & 49.2 & 10.5 & $12 + 1 + 14 + 6 = 33$ \\
\hline
$(8,5,3)_F + (8,5,-3)_F$ & 0.1 & 2.4 & $11 + 1 + 2 + 3 = 17$ \\
\hline
$(8,3,1)_S$ & 110.4 & 10.6 & $10 + 0 + 14 + 3 = 27$ \\
\hline
\end{tabular}
\end{table}

The benchmark data in Table~\ref{tab:model_benchmarks} illustrate the general scaling pattern discussed above. Models built from singlets, triplets, or other comparatively small electroweak representations are typically constructed and expanded within a few seconds, whereas the cost increases substantially for larger non-Abelian multiplets and for field contents admitting many independent invariant contractions. The table also makes clear that the dominant difficulty is associated primarily with the scalar sector rather than with the fermionic one, since scalars admit cubic and quartic self-interactions and therefore a much larger set of candidate invariant operators, whereas fermions do not generate analogous cubic or quartic self-couplings at the renormalizable level. This trend is especially visible in the scalar sector, where repeated large representations can generate a sizable scalar potential and therefore a more expensive symbolic construction stage. It should be emphasized that several of the comparatively fast benchmark points involving large real weak multiplets or color octets benefit directly from the representation-aware optimizations described above; without these specialized constructions, their runtimes would be significantly larger.

\subsection{Expansion into Component Fields}

\paragraph{Motivation.}
A compact gauge-invariant operator basis is useful for classification, but phenomenological calculations often require explicit expressions in terms of component fields. The package therefore provides a deterministic expansion backend that maps every supported scalar, Yukawa, and fermion-mass operator to its corresponding sum of component-field terms.

\paragraph{Algorithmic Structure.}
The component-expansion module acts directly on the gauge-invariant operator object. For scalar operators, it constructs the relevant $SU(2)$ and $SU(3)$ singlet structures and enumerates the contributing weak-isospin and color assignments. The implementation employs explicit Clebsch--Gordan coefficients for weak contractions together with an explicit color-singlet basis. Each resulting term is assigned the appropriate numerical or symbolic coefficient, including the phase conventions associated with the chosen basis.

For Yukawa and fermion-mass operators, the same procedure is applied while accounting for fermion ordering, possible dual-basis slots, and the phase conventions required to maintain consistency with the internal symbolic representation. It is important to note that, in all Yukawa terms, family/generation indices are suppressed.

Mathematically, the expansion is implemented as a multilinear-algebra problem in the representation spaces associated with the gauge quantum numbers of the fields. Each operator is viewed as a tensor product of field factors carrying weak-isospin and color indices, and the backend constructs the corresponding gauge-singlet tensors by combining the relevant Clebsch--Gordan structures for $SU(2)_L$ with explicit invariant tensors for $SU(3)_c$. The component expressions are then obtained by contracting these invariant tensors with the field-component basis in a fixed convention. In this way, the expanded terms are derived directly from the representation theory of the gauge group rather than from hand-written case-by-case rules.

The resulting output is not a heuristic rewriting of the operator name but an explicit decomposition of the gauge singlets into their field components. This representation is subsequently reused by the mass-matrix extractor, avoiding the need for a separate EWSB expansion stage.

\paragraph{Compact and Fully Expanded Reports.}
The report backend exposes both a compact and a fully expanded representation of the Lagrangian. The compact form preserves the gauge-invariant operator notation, whereas the expanded form writes each supported operator class explicitly in terms of component fields. The current \LaTeX{} report output organizes the material by sector, namely Yukawa and fermion-mass terms, the scalar potential, and gauge/kinetic terms.

\subsection{Electroweak Symmetry Breaking and Mass Matrices}

\paragraph{EWSB Workflow.}
The package supports an EWSB reporting stage in which all scalar components that are simultaneously color neutral and electrically neutral are shifted by their corresponding VEVs, and the resulting tree-level mass matrices are derived. In the Standard Model this reduces to the usual Higgs neutral component, while in BSM models the same procedure is applied automatically to every additional neutral color-singlet scalar component. For each such field, the component-level shift takes the form
\begin{align}
\phi_i^0 &\rightarrow \phi_i^0 + \frac{v_i}{\sqrt{2}}, \quad
{\phi_i^0}^{\dagger} \rightarrow {\phi_i^0}^{\dagger} + \frac{v_i}{\sqrt{2}}.
\end{align}
This shift is inserted directly at the component level rather than replacing the field by its vacuum value only. All VEVs are treated as real parameters. 

\paragraph{Stationary Conditions.}
From the scalar potential, the package constructs the vacuum potential by replacing the shifted neutral components by their VEV parts and summing the resulting terms, including Hermitian conjugates when appropriate. The stationary condition is then written as
\begin{align}
\frac{\partial V}{\partial v_i} = 0,
\end{align}
for every VEV symbol $v_i$ associated with a color- and charge-neutral scalar component. In the simplest Standard Model-like configuration this reduces to the Higgs tadpole equation, while in extended models the package reports one stationary condition for each independent neutral VEV generated by the field content.

\paragraph{Mass-Matrix Extraction.} For 
the mass-matrix extraction,  the VEV substitutions are applied term by term. The resulting expressions are scanned for contributions with exactly two surviving fields. These bilinears define the quadratic part of the shifted Lagrangian and are accumulated into symbolic matrices.

For general scalar sectors, the backend stores  hermitian blocks of the form $\phi_i^{\dagger}\phi_j$. This is a natural representation in the original complex component basis. However, neutral scalar fields are presented in a combined real basis for readability. If the model contains neutral complex fields $H^0,\phi_1^0,\phi_2^0,\dots$, the report constructs the neutral scalar mass matrix in the basis
\begin{align}
\big(\mathrm{Re}\,H^0,\,\mathrm{Re}\,\phi_1^0,\,\mathrm{Re}\,\phi_2^0,\dots,\mathrm{Im}\,H^0,\,\mathrm{Im}\,\phi_1^0,\,\mathrm{Im}\,\phi_2^0,\dots\big).
\end{align}
If a neutral field is real, it contributes only to the real part of the basis. Charged or otherwise non-neutral scalar sectors remain listed in their residual hermitian and holomorphic blocks.

The fermion mass matrices are obtained analogously by collecting the bilinears left after the VEV shifts in the Yukawa and fermion-mass sectors. The entire procedure is therefore algorithmic and model independent within the renormalizable scope of the code.

\section{LLM-Assisted Interaction Layer} \label{sec:04}
\subsection{Tool-Calling Design}
The conversational layer is built around a tool-calling interface. The symbolic backend exposes a small number of high-level actions, most importantly a model-construction tool and a report-generation tool. The LLM receives a system prompt describing when to call these tools and is expected to return either plain text or a structured tool invocation with a machine-readable argument payload. In practical terms, the LLM is responsible for tasks such as
\begin{enumerate}[leftmargin=*]
\item interpreting a natural-language description of BSM field content,
\item translating that description into the structured field schema expected by the backend,
\item asking follow-up questions or triggering a confirmation stage if the user input is ambiguous,
\item selecting whether the user is requesting model construction, report generation, or a follow-up action on the current model,
\item formatting the deterministic tool output into a concise human-readable answer.
\end{enumerate}
What the LLM does \emph{not} do is equally important: it does not derive the gauge singlets, compute anomaly coefficients, expand the operators, or calculate the mass matrices. Those tasks remain entirely within the backend code.

\subsection{Conversation State and Confirmation Workflow}
The chat controller maintains a conversation history, the most recent current-model state, and any pending confirmation request. If the user supplies an incomplete field description, the system can infer a candidate structured interpretation and then explicitly ask for confirmation before creating the model. This guardrail is particularly useful in BSM model building, where a missing statement about whether a field is scalar or fermionic, real or complex, or the sign of a hypercharge can materially change the generated theory. More specifically, the user can define a new field through natural-language input by specifying only:
\begin{itemize}
\item[$\magentadiamond$] its quantum numbers under the SM gauge group;

\item[$\magentadiamond$] its spin assignment (scalar or fermion);

\item[$\magentadiamond$] whether the scalar is real, if applicable.
\end{itemize}

Once the user confirms the interpretation, the controller calls the backend tool, stores the resulting structured state, and allows subsequent requests such as report generation or EWSB analysis to operate on the same model context.

After a model has been constructed, the same state mechanism also supports resetting the active model context. The user may explicitly ask the package to clear the current model, in which case the controller discards the stored model state and returns the conversation to a fresh construction state. This allows the user to start over and define a different BSM model without residual fields, assumptions, or analysis results from the previous model carrying over.

\subsection{Supported LLM Backends}

The provider layer is backend agnostic and separates the symbolic workflow from the underlying language model implementation. The package supports three provider classes: local Ollama inference, remote self-hosted model servers accessed through the implemented remote provider interface, and commercial hosted APIs via OpenAI and Anthropic.

\begin{enumerate}[leftmargin=*]
\item \textbf{Local self-hosted models.} The model is served on the same machine as the symbolic package, typically through Ollama. In this mode, the user must first pull a compatible local model, for example,
\begin{align*}
\texttt{ollama pull gpt-oss:20b}
\end{align*}
and then invoke the installed CLI with the corresponding model name; for example,
\begin{align*}
\texttt{bsm-agent --model gpt-oss:20b}
\end{align*}

\item \textbf{Remote self-hosted models.} The model is served on a user-managed remote server and accessed through the implemented remote provider interface. In the current implementation, this corresponds to a remote self-hosted deployment configured through the appropriate provider-prefixed model target together with the remote server settings; for example,
\begin{align*}
\texttt{bsm-agent --model openai:qwen3.6:35b-a3b --api-base <server-url> --api-key <token>}
\end{align*}

\item \textbf{Commercial API models.} The framework can also connect to third-party hosted language models through commercial provider APIs, in particular OpenAI and Anthropic endpoints. Backend-specific wrappers translate the internal conversation history and tool schema into the format required by the selected service. For example, an OpenAI model may be accessed through
\begin{align*}
\texttt{bsm-agent --model openai:gpt-4.1 --api-key <token>}
\end{align*}
\end{enumerate}

Because the orchestration logic is independent of the provider implementation, the symbolic model-building workflow remains unchanged across backends. Users may therefore switch between local self-hosted models, remote self-hosted deployments, and commercial API services without modifying the symbolic model-construction code.

\subsection{Software requirements}
The software requirements of the current implementation are modest. The package metadata explicitly requires \texttt{Python} $\geq 3.10$ and \texttt{setuptools} $\geq 61$ for installation, while the symbolic backend depends on \texttt{SymPy}~\cite{Meurer2017SymPy}. Since the intended user workflow is through the conversational interface with local Ollama-hosted models, the practical software stack also includes \texttt{langchain-core}~\cite{Chase_LangChain_2022}, \texttt{langchain-ollama}, and a local \texttt{ollama}~\cite{Ollama2023} installation.  For local inference, the user must additionally download a compatible Ollama model, for example, \texttt{gpt-oss:20b} or \texttt{qwen3.5:35b}. Alternatively, the same interface can be connected either to a remote self-hosted model server through the implemented remote provider interface or to commercial hosted APIs through the OpenAI or Anthropic provider configurations. PDF report generation further requires a local \texttt{pdflatex} installation. Support for these additional provider classes is implemented in the source code through the corresponding backend wrappers, although the Ollama-based workflow remains the primary intended mode of use.

\subsection{Package installation}\label{installation}

In this work, we demonstrate the installation and usage of the package in a Linux-based operating system environment. The examples presented here were tested on Ubuntu 24.04 running on a Dell Precision 5690 workstation laptop equipped with an Intel\textsuperscript{\textregistered} Core\texttrademark~Ultra 9 185H processor and 64\,GiB of memory.

The package is available at \href{https://github.com/saad-hep/bsm_agent}{\faGithub~bsm_agent}. For example, suppose the user has downloaded and stored it in the ``Downloads'' directory. To install the package, navigate to the corresponding folder from the terminal as follows:
\begin{terminal}
\begin{verbatim}
cd Downloads/bsm_agent
\end{verbatim}
\end{terminal}

Then install the package using the following command:
\begin{terminal}
\begin{verbatim}
pip install .
\end{verbatim}
\end{terminal}

After the installation is complete, the following message should be displayed:
\begin{terminal}
\begin{verbatim}
Successfully installed bsm-agent-0.1.0
\end{verbatim}
\end{terminal}

Before running the package, the user must first download an LLM model. For example, the free \texttt{gpt-oss:20b} model can be downloaded using:
\begin{terminal}
\begin{verbatim}
ollama pull gpt-oss:20b
\end{verbatim}
\end{terminal}
As mentioned above, instead of using a GPT-family model, the user may choose other LLMs, such as models from the Qwen family.

After the model has been downloaded, the package can be launched using:
\begin{terminal}
\begin{verbatim}
bsm-agent --model gpt-oss:20b
\end{verbatim}
\end{terminal}
This command loads the package and initializes the selected LLM. Upon successful startup, the following messages should be displayed:
\begin{terminal}
\begin{verbatim}
[INFO] Loading local model `gpt-oss:20b' via Ollama...
[INFO] BSM agent ready.
\end{verbatim}
\end{terminal}
The package is now ready to assist in building and analyzing BSM models.

\subsection{Installation and usage on macOS and Windows}
The installation procedure on macOS and Windows follows the same general pattern as on Ubuntu. Here, it is assumed that the repository directory is under the user's Downloads folder, matching the Ubuntu/Linux-based example as above. It should be clarified that the author has not verified these installation and usage procedures on Windows or macOS. Testing was conducted exclusively on a Linux-based system, where the packages executed seamlessly.

The user first creates a Python environment with \texttt{Python} $\geq 3.10$, installs the package from the repository root with \texttt{pip install .}, and then launches the installed command-line entry point \texttt{bsm-agent}. In all cases, local usage with Ollama additionally requires that \texttt{ollama} itself be installed on the system, that the desired model already be downloaded, and that the corresponding executable be accessible through the system \texttt{PATH}. If PDF generation is requested, a working \texttt{pdflatex} installation must also be available.

On macOS, one may use the following sequence of commands:
\begin{macOS}
\begin{verbatim}
cd ~/Downloads/bsm_agent
python3 -m venv .venv
source .venv/bin/activate
python3 -m pip install --upgrade pip
python3 -m pip install .
ollama pull gpt-oss:20b
bsm-agent --model gpt-oss:20b
\end{verbatim}
\end{macOS}

On Windows, the same workflow can be executed from PowerShell as:
\begin{Windows}
\begin{verbatim}
cd $HOME\Downloads\bsm_agent
py -3.10 -m venv .venv
.venv\Scripts\Activate.ps1
py -m pip install --upgrade pip
py -m pip install .
ollama pull gpt-oss:20b
bsm-agent --model gpt-oss:20b
\end{verbatim}
\end{Windows}

Alternatively, from the traditional Windows Command Prompt one may use:
\begin{Windows}
\begin{verbatim}
cd %USERPROFILE%\Downloads\bsm_agent
py -3.10 -m venv .venv
.venv\Scripts\activate.bat
py -m pip install --upgrade pip
py -m pip install .
ollama pull gpt-oss:20b
bsm-agent --model gpt-oss:20b
\end{verbatim}
\end{Windows}

If the shell does not recognize the installed \texttt{bsm-agent} command immediately, the same interface can be started explicitly through Python, for example with \texttt{python3 -m bsm\_agent.cli --model gpt-oss:20b} on macOS or \texttt{py -m bsm\_agent.cli --model gpt-oss:20b} on Windows. Likewise, the selected model name need not be \texttt{gpt-oss:20b}; any compatible local Ollama model or appropriately configured remote backend supported by the package may be used instead.

\section{The Standard Model}\label{sec:05}

\subsection{Generating the SM report}
Before constructing new models, we first examine the Standard Model. This example illustrates how the package can be used to analyze the one and only existing model and generate a corresponding report. The user may enter the following natural-language request:
\begin{user}
generate pdf for Standard Model
\end{user}
The package will then generate the corresponding report:
\begin{agent}
PDF generated:
/home/user/Downloads/bsm_agent/output/SM_report.pdf
\end{agent}

Although we will not discuss the contents of the generated PDF in detail, it is useful to highlight some of its basic features. The complete SM report is available in the GitHub repository:
\href{https://github.com/saad-hep/bsm_agent}{\faGithub~bsm_agent}.

The particle content listed in the generated PDF is shown below. As expected, the report includes all Standard Model fermions and the Higgs boson, together with their corresponding quantum numbers.  Gauge bosons ($B_\mu, W_\mu, G_\mu$), however, are not listed.

\begin{pdf}
\[
\begin{aligned}
q & =  (3,2,1/6) \\
l & = (1,2,-1/2) \\
d^{C} & = (\overline{3},1,1/3) \\
u^{C} & = (\overline{3},1,-2/3) \\
e^{C} & =  (1,1,1) \\
H & =  (1,2,1/2)
\end{aligned}
\]
\end{pdf}
The package suppresses generation (family) indices when presenting the particle content. Consequently, only a single representative of each fermion multiplet is displayed. The package adopts the following notation for the components of the electroweak doublets:
\[
\begin{aligned}
q & =  \begin{pmatrix}
    q^{2/3} \\ q^{-1/3}
\end{pmatrix}, \quad 
l & = \begin{pmatrix}
    l^{0} \\ l^{-}
\end{pmatrix}, \quad
H & = \begin{pmatrix}
    H^{+} \\ H^{0}
\end{pmatrix},
\end{aligned}
\]
with the obvious identifications
\[
\begin{aligned}
q\equiv Q, \quad l\equiv L,\quad 
q^{2/3}_i\equiv u_i,\quad  q^{-1/3}_i\equiv d_i, \quad   l^{0}_i\equiv \nu_i, \quad l^{-}_i\equiv e^-_i,
\end{aligned}
\]
with $i=1,2,3$ is the family index.

All Yukawa interactions are listed in Section~2.1. 
\begin{pdf}
Yukawa terms:
\begin{dmath*}
Y_{1} \, H^\dagger d^{C} q + Y_{2} \, u^{C} q H + Y_{3} \, H^\dagger e^{C} l+  \mathrm{h.c.}
\end{dmath*}
\end{pdf}
As mentioned above, family (as well as Lorentz) indices are always suppressed throughout the report.  For example, the first Yukawa coupling shown above must be interpreted as
\begin{align*}
Y_{1} \, H^\dagger d^C q= Y^{ij}_{1} \, H^\dagger d^C_i q_j, \quad i,j= 1,2,3.
\end{align*}
More specifically, the Lorentz invariant term is constructed as:
\begin{dmath*}
Y^{ij}_{1} \, H^\dagger {d^{\alpha C}_{i}} q_{j}^{\beta} \epsilon_{\alpha\beta},
\end{dmath*}
where $\alpha, \beta$ are spinor indices.

All the pure scalar interactions are listed in Section 2.2. 
\begin{pdf}
Scalar Potential:
\begin{dmath*}
m^{2}_{1} \, H H^\dagger +
\lambda_{2} \, H H H^\dagger H^\dagger
\end{dmath*}
\end{pdf}
Finally, the gauge kinetic terms are listed in Section 2.3. 
\begin{pdf}
\begin{dmath*}
-\frac{1}{4} B_{\mu\nu} B^{\mu\nu}
-\frac{1}{4} W^a_{\mu\nu} W^{a,\mu\nu}
-\frac{1}{4} G^A_{\mu\nu} G^{A,\mu\nu}
+
(D_\mu q)^\dagger D^\mu q
+
(D_\mu l)^\dagger D^\mu l
+
(D_\mu d^{C})^\dagger D^\mu d^{C}
+
(D_\mu u^{C})^\dagger D^\mu u^{C}
+
(D_\mu e^{C})^\dagger D^\mu e^{C}
+
(D_\mu H)^\dagger D^\mu H
\end{dmath*}
\end{pdf}

\subsection{Expanding the SM Lagrangian} 
 One of the most powerful capabilities of the package is the automatic expansion of Lagrangian terms into their component-field expressions. When requested, the package fully expands all $SU(3)_c$ and $SU(2)_L$ group indices, yielding explicit component-level interactions. A key advantage is that the user is not required to define the component fields manually. Instead, the package automatically determines the appropriate decomposition from the gauge structure and quantum-number assignments of the fields.
\begin{user}
    expand Lagrangian
\end{user}
\begin{agent}
\begin{verbatim}
Generated report for 'SM' with field content SM. 
TeX: /home/user/Downloads/bsm_agent/output/SM_report.tex. 
PDF: /home/user/Downloads/bsm_agent/output/SM_report.pdf. 
[2.5s]    
\end{verbatim}    
\end{agent}

\begin{pdf}
Expanded Yukawa interactions:
\begin{dmath*}
Y_{1}\left( \begin{array}{l}{{H}^{+}}^\dagger {d^{C}}^{+1/3,\bar{b}} {q}^{+2/3}_{b} + {{H}^{+}}^\dagger {d^{C}}^{+1/3,\bar{g}} {q}^{+2/3}_{g} \\\\ + {{H}^{+}}^\dagger {d^{C}}^{+1/3,\bar{r}} {q}^{+2/3}_{r} + {{H}^{0}}^\dagger {d^{C}}^{+1/3,\bar{b}} {q}^{-1/3}_{b} \\\\ + {{H}^{0}}^\dagger {d^{C}}^{+1/3,\bar{g}} {q}^{-1/3}_{g} + {{H}^{0}}^\dagger {d^{C}}^{+1/3,\bar{r}} {q}^{-1/3}_{r}\end{array} \right) + \mathrm{h.c.}
\end{dmath*}

\begin{dmath*}
Y_{2}\left( \begin{array}{l}{u^{C}}^{-2/3,\bar{b}} {q}^{+2/3}_{b} {H}^{0} - {u^{C}}^{-2/3,\bar{b}} {q}^{-1/3}_{b} {H}^{+} \\\\ + {u^{C}}^{-2/3,\bar{g}} {q}^{+2/3}_{g} {H}^{0} - {u^{C}}^{-2/3,\bar{g}} {q}^{-1/3}_{g} {H}^{+} \\\\ + {u^{C}}^{-2/3,\bar{r}} {q}^{+2/3}_{r} {H}^{0} - {u^{C}}^{-2/3,\bar{r}} {q}^{-1/3}_{r} {H}^{+}\end{array} \right) + \mathrm{h.c.}
\end{dmath*}

\begin{dmath*}
Y_{3}\left( {{H}^{+}}^\dagger {e^{C}}^{+} {l}^{0} + {{H}^{0}}^\dagger {e^{C}}^{+} {l}^{-} \right) + \mathrm{h.c.}
\end{dmath*}    
\end{pdf}

\begin{pdf}
Expanded scalar potential:
\begin{dmath*}
m^{2}_{1}\left( {H}^{0} {{H}^{0}}^\dagger + {H}^{+} {{H}^{+}}^\dagger \right)
\end{dmath*}

\begin{dmath*}
\lambda_{2}\left( \begin{array}{l}{H}^{0} {H}^{0} {{H}^{0}}^\dagger {{H}^{0}}^\dagger + 2\, {H}^{0} {H}^{+} {{H}^{+}}^\dagger {{H}^{0}}^\dagger  + {H}^{+} {H}^{+} {{H}^{+}}^\dagger {{H}^{+}}^\dagger\end{array} \right)
\end{dmath*}
\end{pdf}
The first term with coefficient $m^2_1$ shows that the Higgs field is canonically normalized, as expected. For any other term in the scalar potential, the package does not put any extra numerical factors with the coefficients.

We do not present the expanded gauge kinetic terms here. However, we point out that the expanded form of the gauge kinetic terms, even in the SM, is rather lengthy.

\subsection{Generating the Mass matrices via EW symmetry breaking} 
The user may request the mass matrices using the following natural-language prompt:
\begin{user}
    generate mass matrix
\end{user}

\begin{agent}
\begin{verbatim}
Generated report for 'SM' with field content SM.
TeX: /home/user/Downloads/bsm_agent/output/SM_report.tex. 
PDF: /home/user/Downloads/bsm_agent/output/SM_report.pdf. 
[2.3s]    
\end{verbatim}    
\end{agent}

As can be seen from the report, the package substitutes the vacuum expectation value (VEV) of the SM Higgs doublet according to
\begin{pdf}
\begin{align*}
&{H}^{0} \to {H}^{0} + \frac{v}{\sqrt{2}},
\quad {{H}^{0}}^\dagger \to {{H}^{0}}^\dagger + \frac{v}{\sqrt{2}}
\end{align*}
\end{pdf}
where $v \simeq 246~\mathrm{GeV}$. The package also provides the corresponding stationary condition:
\begin{pdf}
\begin{align*}
\frac{\partial V}{\partial v} = v \left(\lambda_{2} v^{2} + m^{2}_{1}\right) = 0
\end{align*}
\end{pdf}

Looking carefully at the scalar masses, one finds that, upon imposing the stationary condition
$\lambda_{2} v^{2} + m^{2}_{1}=0$, the expected massless Goldstone modes are correctly recovered:
\begin{pdf}
Neutral Scalar Mass Matrix:
\[
\begin{array}{l}
M^2_{\text{neutral}}[\operatorname{Re}({H}^{0}), \operatorname{Im}({H}^{0});\operatorname{Re}({H}^{0}), \operatorname{Im}({H}^{0})] = 
\left[\begin{matrix}3 \lambda_{2} v^{2} + m^{2}_{1} & 0\\0 & \lambda_{2} v^{2} + m^{2}_{1}\end{matrix}\right]
\end{array}
\]
Scalar Hermitian Blocks:
\[
\begin{array}{l}
M^2_{1}[\left({H}^{+}\right)^\dagger;{H}^{+}] = 
\left[\begin{matrix}\lambda_{2} v^{2} + m^{2}_{1}\end{matrix}\right]
\end{array}
\]
\end{pdf}
The package explicitly specifies the basis in which each mass matrix is written.

Using the stationary condition, the masses of $\operatorname{Im}({H}^{0})$ and ${H}^{+}$ vanish, as expected for the Goldstone modes associated with electroweak symmetry breaking. The remaining physical scalar corresponds to the SM Higgs boson, with mass
\[
m_h^2 = 2\lambda_2 v^2.
\]

Let us now examine the SM fermion mass matrices. For colored fermions, we present only the results corresponding to the red ($r$) color component, since the mass matrices are identical for all color components.
\begin{pdf}
Down-type quark mass matrix:
\[
\begin{array}{l}
M_{3}[{d^{C}}^{+1/3,\bar{r}};{q}^{-1/3}_{r}] = 
\left[\begin{matrix}\frac{\sqrt{2} Y_{1} v}{2}\end{matrix}\right]
\end{array}
\]

Charged lepton mass matrix:
\[
\begin{array}{l}
M_{4}[{e^{C}}^{+};{l}^{-}] = 
\left[\begin{matrix}\frac{\sqrt{2} Y_{3} v}{2}\end{matrix}\right]
\end{array}
\]

Up-type quark mass matrix:
\[
\begin{array}{l}
M_{7}[{u^{C}}^{-2/3,\bar{r}};{q}^{+2/3}_{r}] = 
\left[\begin{matrix}\frac{\sqrt{2} Y_{2} v}{2}\end{matrix}\right]
\end{array}
\]
\end{pdf}
These matrices, as expected, reproduce the well-known SM results. Restoring the generation indices, the mass matrices are written in the bases $d^{C}_i d_j$, $u^{C}_i u_j$, and $e^{C}_i e^-_j$. Accordingly, the corresponding mass matrices are given by
\begin{align*}
m^{ij}_d=  Y_{1}^{ij} \, \frac{v}{\sqrt{2}}, \quad m^{ij}_u=  Y_{2}^{ij} \, \frac{v}{\sqrt{2}}, \quad m^{ij}_e=  Y_{3}^{ij} \, \frac{v}{\sqrt{2}}.
\end{align*}
It is important to note that the package derives all mass matrices in a generic flavor basis. In the case of the SM, the Yukawa coupling matrices $Y_{1,2,3}$ can be diagonalized simultaneously through appropriate field redefinitions. More generally, for any BSM model generated by the package, all mass matrices are presented in a generic flavor basis. The user is therefore free to perform any desired basis transformation and work in the basis most suitable for the analysis at hand.

\section{Building our first BSM model}\label{sec:06}

If desired, the user can check whether a model has already been defined by entering the following natural-language query:
\begin{user}
is there a predefined model?
\end{user}
The package then displays the list of available predefined models and prompts the user to verify that the corresponding quantum numbers have been interpreted correctly.
\begin{agent}
No built‑in ``predefined'' BSM model is available.  
To create a model you need to specify the field content (name, representation, hypercharge, etc.). Once you provide that list, I can build the model and generate a report.
\end{agent}

Let us now construct our first BSM model, which we label as ``Model 1''. As a simple example, we consider an extension of the SM with right-handed neutrinos. Naturally, the construction of any model begins by specifying the quantum numbers of the new particle(s).
\begin{user}
build a model with a fermion singlet
\end{user}
At this stage, the package checks with the user if the quantum numbers are correctly interpreted. 
\begin{agent}
I interpreted the BSM field content as fermion (1,1,0). If this is correct, type only yes. If not, restate the quantum numbers.
\end{agent}

\begin{user}
yes
\end{user}

\begin{agent}
The model is constructed. [0.1s]
\end{agent}

As simple as that, the model has been constructed.

The user may then ask the package to describe the basic properties of the newly constructed model:
\begin{user}
describe the model
\end{user}
In such cases, the package provides a brief summary of the model. First, it reports the quantum numbers of the particle(s) added to the SM. Second, it summarizes the structure of the Lagrangian by counting the generated Yukawa interactions, mass terms, scalar-potential terms, and gauge kinetic terms. Finally, it determines whether the constructed model is anomaly-free.
\begin{agent}
\begin{verbatim}
Model: SM+psi_1
Field content:
- psi_1 is a fermion in (1,1,0): color singlet, weak singlet, hypercharge 0.

Interactions and consistency:
  - Kinetic terms: 10
  - Fermion mass: 1
  - Scalar potential: 2
  - Yukawa: 4
  - Mixed BSM interaction terms: 0
  - Status: anomaly-free
  - Anomaly summary: anomaly-free: su3_cubic=0, su2_su2_u1=0, su3_su3_u1=0, 
    u1_gravity=0, u1_cubic=0
\end{verbatim}
\end{agent}
From this summary, we can quickly see that we have constructed a viable model by looking at the ``Status'': \textbf{Status: \textcolor{darkergreen}{anomaly-free}}.

Let us go ahead and request the package to generate a PDF report.
\begin{user}
generate pdf
\end{user}
\begin{agent}
\begin{verbatim}
Generated report for 'SM+psi_1' with field content (1,1,0). 
TeX: /home/user/Downloads/bsm_agent/output/SM_psi_1.tex. 
PDF: /home/user/Downloads/bsm_agent/output/SM_psi_1.pdf. 
[0.7s]   
\end{verbatim}
\end{agent}
Let us now briefly examine the generated PDF. In general, the report is organized into two main sections.
\begin{pdf}
\centering
\includegraphics[width=0.65\textwidth]{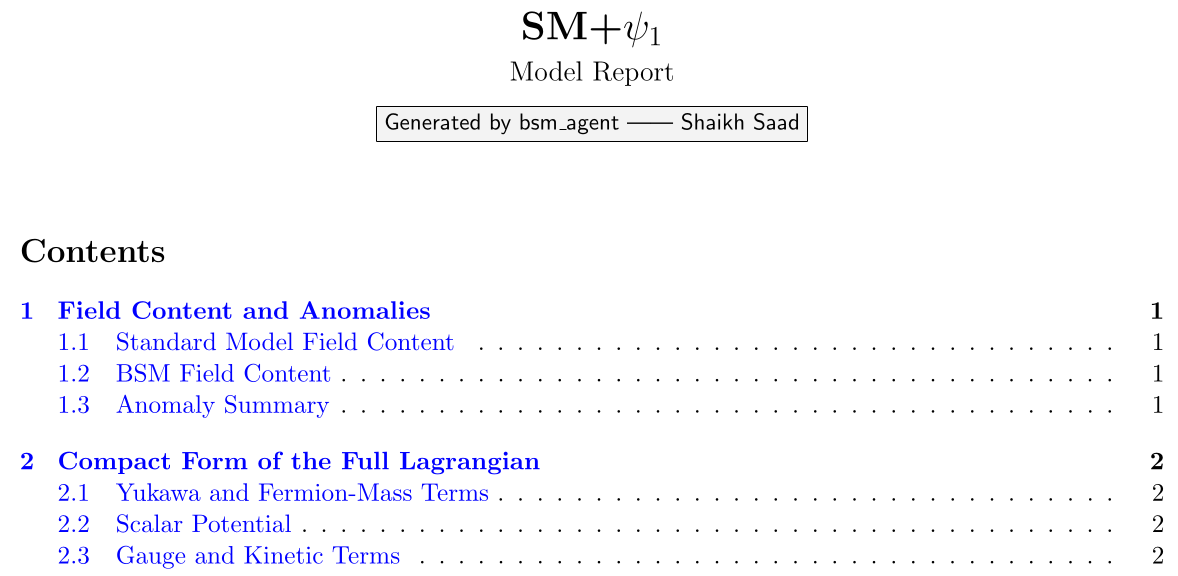}
\end{pdf}

Section~1 summarizes the field content of the model and reports its gauge-anomaly status.
\begin{pdf}
BSM Field Content:
\[
\begin{aligned}
\textcolor{red}{\psi_{1}} & = & \text{fermion} & \quad (1,1,0)
\end{aligned}
\]

Anomaly Summary:\\
\textbf{Anomalies:} \textcolor{green!50!black}{\textbf{free}}.
\[
\begin{array}{@{} l c @{}}
\text{Channel} & \text{Value} \\
SU(3)c-SU(3)c-SU(3)c & 0 \\
SU(2)L-SU(2)L-U(1)Y & 0 \\
SU(3)c-SU(3)c-U(1)Y & 0 \\
U(1)Y-gravity & 0 \\
U(1)Y-U(1)Y-U(1)Y & 0 \\
\end{array}
\]
\end{pdf}

Section~2 presents the full Lagrangian in a compact form. For convenience, it is divided into three subsections: (i) Yukawa interactions, (ii) the scalar potential, and (iii) gauge kinetic terms.

On top of the SM Yukawa couplings (represented below by ellipses for brevity; the package lists all terms explicitly), the report also includes the BSM Yukawa interactions.
\begin{pdf}
Yukawa Terms:
\begin{dmath*}
... + Y_{4} \, \textcolor{red}{\psi_{1}} l H + 
M_{1} \, \textcolor{red}{\psi_{1}} \textcolor{red}{\psi_{1}} + \mathrm{h.c.}
\end{dmath*}
\end{pdf}
The new physics part of the Yukawa interactions:
\begin{dmath*}
Y_{4} \, \textcolor{red}{\psi_{1}} l H + 
M_{1} \, \textcolor{red}{\psi_{1}} \textcolor{red}{\psi_{1}},
\end{dmath*}
must be interpreted as
\begin{dmath*}
Y_{4}^{Ij} \, \textcolor{red}{\psi_{1}}_I l_j H + 
M_{1}^{IJ} \, \textcolor{red}{\psi_{1}}_I \textcolor{red}{\psi_{1}}_J, \quad I,J= 1, ..., N.
\end{dmath*}

Since the scalar potential is identical to that of the SM, we do not reproduce it here. Finally, the new-physics contribution to the kinetic terms is given by:
\begin{pdf}
Gauge and Kinetic Terms:
\begin{dmath*}
...+(D_\mu \textcolor{red}{\psi_{1}})^\dagger D^\mu \textcolor{red}{\psi_{1}}
\end{dmath*}
\end{pdf}

\section{Package features}\label{sec:07}
In this section, our goal is not to construct phenomenologically viable models. Rather, we focus on illustrating various aspects of the package's notation and model-building interface. The primary purpose of this section is to demonstrate the flexibility afforded by the chat-based construction framework: users are not required to follow any rigid syntax or predefined input format. Instead, models can be specified using natural-language descriptions, which is one of the most powerful features of the package.

\subsection{Building another model}
Suppose the user has constructed the BSM model described in Sec.~\ref{sec:06}. Before building a new model, the previously defined model must be removed. This can be accomplished in several ways. For example, the user may exit the package (by pressing \texttt{Ctrl+D}) and then restart it (with \texttt{bsm-agent --model gpt-oss:20b}).

Alternatively, the user may instruct the package to clear the current model using the following natural-language request:
\begin{user}
clear
\end{user}

\begin{agent}
    Conversation cleared.
\end{agent}
or, 
\begin{user}
remove the model
\end{user}

\begin{agent}
    Model removed.
\end{agent}

or, 
\begin{user}
reset model
\end{user}

\begin{agent}
   Cleared the current model.
\end{agent}

or, 
\begin{user}
start over
\end{user}
etc. 

After clearing the model, the user can verify that no model is currently loaded by issuing the following natural-language request:
\begin{user}
summary
\end{user}
\begin{agent}
No model is currently defined.
\end{agent}

\subsection{Extending an Existing Model}
Let us again consider the model constructed in Sec.~\ref{sec:06}. This model, which we refer to as Model~1, contains $N$ generations of a fermionic singlet with quantum numbers $(1,1,0)$, denoted by ${\color{red}\psi_i}$.

Suppose the user wishes to \textbf{extend} this model further by introducing an additional BSM field, namely a scalar singlet. Rather than constructing a new model from scratch, the package allows the user to build directly upon the existing model.

\begin{user}
add a scalar (1,1,0)
\end{user}

\begin{agent}
    I interpreted the BSM field content as scalar (1,1,0). If this is correct, type only yes. If not, restate the quantum numbers.
\end{agent}

\begin{user}
YES
\end{user}

\begin{agent}
The model is constructed. [0.1s]
\end{agent}

\begin{user}
summarize
\end{user}

\begin{agent}
\begin{verbatim}
BSM_Agent> Model: SM+psi_1+phi_1
Field content:
- psi_1 is a fermion in (1,1,0): color singlet, weak singlet, hypercharge 0.
- phi_1 is a scalar in (1,1,0): color singlet, weak singlet, hypercharge 0.

Interactions and consistency:
  - Kinetic terms: 11
  - Fermion mass: 1
  - Scalar potential: 13
  - Yukawa: 6
  - Mixed BSM interaction terms: 2
  - Status: anomaly-free
  - Anomaly summary: anomaly-free: su3_cubic=0, su2_su2_u1=0, su3_su3_u1=0, 
    u1_gravity=0, u1_cubic=0
\end{verbatim}
\end{agent}
In this manner, the user may incrementally extend an existing model by adding additional fields.

\subsection{Scalar/Fermion field}
When a particle is added to the SM, if the user does not specify whether the new field is a scalar or a fermion, the package automatically requests this information.
\begin{user}
  let us build a model with (8,2,1/2) 
\end{user}
\begin{agent}
I interpreted the BSM field content as (8,2,1/2). You did not specify whether the new field is a scalar or fermion. Reply with either 'scalar' or 'fermion' if these quantum numbers are correct, or restate the full field content.    
\end{agent}
\begin{user}
   scalar
\end{user}
\begin{agent}
    The model is constructed. [48.2s]
\end{agent}
Note that, owing to the more complicated structure of the scalar potential for this field, the construction of the model takes somewhat longer. Nevertheless, the entire process still completed in less than a minute.

\begin{user}
   can you please describe the model
\end{user}

\begin{agent}
\begin{verbatim}
Model: SM+phi_1
Field content:
- phi_1 is a scalar in (8,2,1/2): color octet, weak doublet, hypercharge 1/2.

Interactions and consistency:
  - Kinetic terms: 10
  - Scalar potential: 14
  - Yukawa: 5
  - Mixed BSM interaction terms: 0
  - Status: anomaly-free
  - Anomaly summary: anomaly-free: su3_cubic=0, su2_su2_u1=0, su3_su3_u1=0,
    u1_gravity=0, u1_cubic=0    
\end{verbatim}   
\end{agent}
From this model summary, one can infer that the constructed model contains two new Yukawa interaction terms and twelve new terms in the scalar potential. By explicitly expanding the full Lagrangian, we have verified that the resulting interactions agree with those presented in Ref.~\cite{Manohar:2006ga}, after performing the appropriate mapping to account for the user-defined choice of basis.

\subsection{Real/Complex scalar}
Recall, in Sec. we have constructed a model with a singlet scalar: 
\begin{user}
    build a model with scalar (1,1,0)
\end{user}

\begin{agent}
    I interpreted the BSM field content as scalar (1,1,0). If this is correct, type only yes. If not, restate the quantum numbers.
\end{agent}

\begin{user}
    yes
\end{user}

\begin{agent}
    The model is constructed. [0.1s]
\end{agent}

\begin{user}
    summary
\end{user}

\begin{agent}
\begin{verbatim}
...
  - Scalar potential: 13
...
\end{verbatim}
\end{agent}

\begin{user}
    pdf
\end{user}
 
\begin{pdf}

\begin{align*}
&
\bigg\{\xi^{3}_{1} \, \textcolor{blue}{\phi_{1}} + \mathrm{h.c.}\bigg\}
+
m^{2}_{2} \, H H^\dagger
+
\bigg\{
m^{2}_{3} \, \textcolor{blue}{\phi_{1}} \textcolor{blue}{\phi_{1}} + \mathrm{h.c.}
\bigg\}
+
m^{2}_{4} \, \textcolor{blue}{\phi_{1}} \textcolor{blue}{\phi_{1}}^\dagger
\\&
+
\bigg\{
\mu_{5} \, H H^\dagger \textcolor{blue}{\phi_{1}} 
+
\mu_{6} \, \textcolor{blue}{\phi_{1}} \textcolor{blue}{\phi_{1}} \textcolor{blue}{\phi_{1}} 
+
\mu_{7} \, \textcolor{blue}{\phi_{1}} \textcolor{blue}{\phi_{1}} \textcolor{blue}{\phi_{1}}^\dagger 
+ \mathrm{h.c.}
\bigg\}
+
\lambda_{8} \, H H H^\dagger H^\dagger
+
\bigg\{
\lambda_{9} \, H H^\dagger \textcolor{blue}{\phi_{1}} \textcolor{blue}{\phi_{1}} + \mathrm{h.c.}
\bigg\}
\\&
+
\lambda_{10} \, H H^\dagger \textcolor{blue}{\phi_{1}} \textcolor{blue}{\phi_{1}}^\dagger
+
\bigg\{
\lambda_{11} \, \textcolor{blue}{\phi_{1}} \textcolor{blue}{\phi_{1}} \textcolor{blue}{\phi_{1}} \textcolor{blue}{\phi_{1}} 
+ 
\lambda_{12} \, \textcolor{blue}{\phi_{1}} \textcolor{blue}{\phi_{1}} \textcolor{blue}{\phi_{1}} \textcolor{blue}{\phi_{1}}^\dagger + \mathrm{h.c.}
\bigg\}
+
\lambda_{13} \, \textcolor{blue}{\phi_{1}} \textcolor{blue}{\phi_{1}} \textcolor{blue}{\phi_{1}}^\dagger \textcolor{blue}{\phi_{1}}^\dagger
\end{align*}
\end{pdf}
Note, in the above model construction, the user has not specified if the singlet scalar is real or complex. In such a scenario,  the package, by default, assumes complex scalars. 

To build a model with real scalar (which requires the field carrying zero hypercharge), the user must specify:
\begin{user}
    build a model having real scalar (1,1,0)
\end{user}

\begin{agent}
    interpreted the BSM field content as scalar (1,1,0). If this is correct, type only yes. If not, restate the quantum numbers.
\end{agent}

\begin{user}
    yes
\end{user}

\begin{agent}
    The model is constructed. [0.1s]
\end{agent}

\begin{user}
    summary
\end{user}

\begin{agent}
\begin{verbatim}
...
  Scalar potential: 8
...
\end{verbatim}
\end{agent}
Note, now, due to reality constraint, the number of terms in the scalar potential is no longer 13, but 8. Let us look at the terms generated by asking the package to build a pdf file:
\begin{user}
    pdf
\end{user}
 
\begin{pdf}
\begin{align*}
&
\xi^{3}_{1} \, \textcolor{blue}{\phi_{1}}
+
m^{2}_{2} \, H H^\dagger
+
m^{2}_{3} \, \textcolor{blue}{\phi_{1}} \textcolor{blue}{\phi_{1}}
+
\mu_{4} \, H H^\dagger \textcolor{blue}{\phi_{1}}
+
\mu_{5} \, \textcolor{blue}{\phi_{1}} \textcolor{blue}{\phi_{1}} \textcolor{blue}{\phi_{1}}
+
\lambda_{6} \, H H H^\dagger H^\dagger
\\&
+
\lambda_{7} \, H H^\dagger \textcolor{blue}{\phi_{1}} \textcolor{blue}{\phi_{1}}
+
\lambda_{8} \, \textcolor{blue}{\phi_{1}} \textcolor{blue}{\phi_{1}} \textcolor{blue}{\phi_{1}} \textcolor{blue}{\phi_{1}}
\end{align*}
\end{pdf}

\subsection{Stating the quantum number}
In this subsection, let us focus on how to specify the quantum numbers of new physics states. Suppose we are trying to build a model that contains a fermion with quantum numbers $(\overline 6,3,-1/3)$. The user can specify this information in various ways:
\begin{user}
construct a model with a fermion (6bar, 3, -1/3)
\end{user}
or,
\begin{user}
a fermion of (bar 6, 3, -1/3)
\end{user}
or,
\begin{user}
add a fermion which is color anti-sextet, weak triplet, and has hypercharge minus one third
\end{user}
or,
\begin{user}
build a model with color anti 6, electroweak triplet, hyper charge minus one third
\end{user}
and so on.  In all these cases, if the package understands the notation, it will, as usual, ask:
\begin{agent}
    I interpreted the BSM field content as fermion (bar6,3,-1/3). If this is correct, type only yes. If not, restate the quantum numbers.
\end{agent}

\begin{user}
yes
\end{user}

\begin{agent}
The model is constructed. [0.1s]
\end{agent}

\begin{user}
describe
\end{user}

\begin{agent}
\begin{verbatim}
Model: Anti6TripletModel
Field content:
  - psi_1 is a fermion in (bar6,3,-1/3): color anti-sextet, weak triplet, 
    hypercharge -1/3.

Interactions and consistency:
  - Kinetic terms: 10
  - Scalar potential: 2
  - Yukawa: 3
  - Mixed BSM interaction terms: 0
  - Status: anomalous
  - Anomaly summary: anomalous: su3_cubic=-21, su2_su2_u1=-4, su3_su3_u1=-5/2, 
    u1_gravity=-6, u1_cubic=-2/3 
  - Hint: Try adding the conjugate fermion (6,3,+1/3) to cancel the anomaly.
\end{verbatim}
\end{agent}
This quick summary tells us that this model is not  viable since: \textbf{Status: \textcolor{red}{anomalous}}. 

In such a case, when a model has gauge anomalies, the package may provide \textbf{hint} on what is the simplest possible way to cancel the anomalies. For a fermion $\psi$, it is likely to suggest adding $\psi^c$ to cancel anomalies.

If the package fails to understand the quantum numbers provided for a particle, it asks the user for clarification. For example,
\begin{agent}
    Please restate the quantum numbers for the BSM field content
\end{agent}
At this stage, the user can correct any mistakes in the quantum numbers and continue with the model building process.

\subsection{Basis Choice, Normalization, and Phase Conventions}

It is important to clarify that when the user asks the package to construct a model,

\begin{itemize}

\item The package uses general group-theoretical rules to construct the full Lagrangian. In doing so, it correctly determines the number of independent terms of each type and lists all of them. This part of the construction depends only on group theory and is therefore free from any arbitrariness.

\item On the other hand, when the user asks the package to \textbf{expand} the Lagrangian, the result necessarily depends on a particular choice of basis, normalization, and phase conventions. Therefore, if the user wishes to compare their results with those reported by the package, an appropriate mapping between conventions must be established. For any meaningful comparison, the user should be aware of the following points.

\item[$\star$] The package always uses canonical normalization for all fields. Consequently, the quadratic term for a scalar field $\phi_i$ is written as $\phi_i \phi_i^\dagger$. If $\phi_i$ carries $SU(3)$ and/or $SU(2)$ charges, each component field within $\phi_i$ is also normalized canonically.

\item[$\star$] When writing terms in the Lagrangian, the package does not introduce any numerical factors in front of the corresponding coupling coefficients. Therefore, if a scalar field is real, such as $(1,1,0)$, the package does not include the conventional factor of $1/2$ multiplying the quadratic term. Similarly, no numerical factors are inserted in front of cubic or quartic interaction terms.

\item[$\star$] When comparing results, the user should be aware that the package may employ phase conventions different from their own. For example, under a rephasing of a component field,
\[
\phi_i^\alpha\to e^{i\varphi_\alpha} \phi_i^\alpha, 
\]
where $\alpha$ labels a component field, the canonical normalization remains unchanged. Although different users may adopt different phase conventions, the physical predictions remain identical.

\item[$\star$] The package lists only independent terms in the Lagrangian; redundant or dependent terms are never included. Consequently, when more than one independent term exists for a given operator structure, comparisons between expanded expressions must be performed with care. For example, suppose the package identifies two independent quartic couplings, $\lambda_{k,c1}$ and $\lambda_{k,c2}$, for a scalar field $\phi_i$. If, in a different convention, the user instead works with couplings $\lambda_k$ and $\lambda_k^\prime$, then the two sets of parameters are related through a linear transformation,
\begin{align*}
\lambda_k &= A_1 \lambda_{k,c1} + B_1 \lambda_{k,c2}, \qquad
\lambda_k^\prime = A_2 \lambda_{k,c1} + B_2 \lambda_{k,c2},
\end{align*}
where $A_i$ and $B_i$ are numerical coefficients.

\end{itemize}

\subsubsection{Example case: Weak triplet}
Let us demonstrate this for a specific model consisting of the following BSM state:
\begin{align*}
\textcolor{blue}{\phi_{1}} = \text{scalar} & \quad (1,3,1)
\end{align*}

\begin{user}
    build a model with scalar (1,3,1)
\end{user}

\begin{agent}
    I interpreted the BSM field content as scalar (1,3,1). If this is correct, type only yes. If not, restate the quantum numbers.
\end{agent}

\begin{user}
    yes
\end{user}

\begin{agent}
    The model is constructed. [0.1s]
\end{agent}

\begin{user}
    pdf
\end{user}

\begin{agent}
\begin{verbatim}
    Generated report for 'SM+phi_1' with field content (1,3,1). 
    TeX: /home/user/Downloads/bsm_agent/output/SM_phi_1.tex. 
    PDF: /home/user/Downloads/bsm_agent/output/SM_phi_1.pdf. 
    [1.2s]    
\end{verbatim}
\end{agent}

\begin{user}
expand Lagrangian    
\end{user}

\begin{agent}
 \begin{verbatim}
  Generated report for 'SM+phi_1' with field content (1,3,1). 
  TeX: /home/user/Downloads/bsm_agent/output/SM_phi_1.tex. 
  PDF: /home/user/Downloads/bsm_agent/output/SM_phi_1.pdf. 
  [2.3s]   
 \end{verbatim}   
\end{agent}

\begin{user}
    mass matrix
\end{user}

\begin{agent}
\begin{verbatim}
Generated report for 'SM+phi_1' with field content (1,3,1). 
TeX: /home/user/Downloads/bsm_agent/output/SM_phi_1.tex. 
PDF: /home/user/Downloads/bsm_agent/output/SM_phi_1.pdf. 
[2.8s]    
\end{verbatim}    
\end{agent}

The new physics interactions generated by the package are as follows. 
\begin{pdf}
Yukawa Terms:
\begin{dmath*}
Y_{3} \, l l \textcolor{blue}{\phi_{1}} + \mathrm{h.c.}
\end{dmath*}

Scalar Potential:
\begin{dmath*}
m^{2}_{2} \, \textcolor{blue}{\phi_{1}} \textcolor{blue}{\phi_{1}}^\dagger
+\bigg\{
\mu_{3} \, H H \textcolor{blue}{\phi_{1}}^\dagger + \mathrm{h.c.}
\bigg\}
+
\lambda_{5,c1} \, H H^\dagger \textcolor{blue}{\phi_{1}} \textcolor{blue}{\phi_{1}}^\dagger
+
\lambda_{5,c2} \, H H^\dagger \textcolor{blue}{\phi_{1}} \textcolor{blue}{\phi_{1}}^\dagger
+
\lambda_{7,c1} \, \textcolor{blue}{\phi_{1}} \textcolor{blue}{\phi_{1}} \textcolor{blue}{\phi_{1}}^\dagger \textcolor{blue}{\phi_{1}}^\dagger
+
\lambda_{7,c2} \, \textcolor{blue}{\phi_{1}} \textcolor{blue}{\phi_{1}} \textcolor{blue}{\phi_{1}}^\dagger \textcolor{blue}{\phi_{1}}^\dagger
\end{dmath*}
\end{pdf}

The corresponding expanded forms are:
\begin{pdf}
Expanded Yukawa:
\begin{dmath*}
Y_{3}\left( \begin{array}{l}{l}^{-} {l}^{-} {\textcolor{blue}{\phi_{1}}}^{++} - \frac{\sqrt{2}}{2}\, {l}^{-} {l}^{0} {\textcolor{blue}{\phi_{1}}}^{+}  - \frac{\sqrt{2}}{2}\, {l}^{0} {l}^{-} {\textcolor{blue}{\phi_{1}}}^{+} + {l}^{0} {l}^{0} {\textcolor{blue}{\phi_{1}}}^{0}\end{array} \right) + \mathrm{h.c.}
\end{dmath*}

Expanded scalar couplings:
\begin{dmath*}
m^{2}_{2}\left( \begin{array}{l}{\textcolor{blue}{\phi_{1}}}^{0} {{\textcolor{blue}{\phi_{1}}}^{0}}^\dagger + {\textcolor{blue}{\phi_{1}}}^{+} {{\textcolor{blue}{\phi_{1}}}^{+}}^\dagger  + {\textcolor{blue}{\phi_{1}}}^{++} {{\textcolor{blue}{\phi_{1}}}^{++}}^\dagger\end{array} \right)
\end{dmath*}

\begin{dmath*}
\mu_{3}\left( \begin{array}{l}{H}^{0} {H}^{0} {{\textcolor{blue}{\phi_{1}}}^{0}}^\dagger + \sqrt{2}\, {H}^{0} {H}^{+} {{\textcolor{blue}{\phi_{1}}}^{+}}^\dagger  + {H}^{+} {H}^{+} {{\textcolor{blue}{\phi_{1}}}^{++}}^\dagger\end{array} \right) + \mathrm{h.c.}
\end{dmath*}

\begin{dmath*}
\lambda_{5,c1}^{(1)}\left( \begin{array}{l}\sqrt{2}\, {H}^{0} {{H}^{+}}^\dagger {\textcolor{blue}{\phi_{1}}}^{+} {{\textcolor{blue}{\phi_{1}}}^{0}}^\dagger + \sqrt{2}\, {H}^{0} {{H}^{+}}^\dagger {\textcolor{blue}{\phi_{1}}}^{++} {{\textcolor{blue}{\phi_{1}}}^{+}}^\dagger \\\\ + {H}^{0} {{H}^{0}}^\dagger {\textcolor{blue}{\phi_{1}}}^{0} {{\textcolor{blue}{\phi_{1}}}^{0}}^\dagger - {H}^{0} {{H}^{0}}^\dagger {\textcolor{blue}{\phi_{1}}}^{++} {{\textcolor{blue}{\phi_{1}}}^{++}}^\dagger \\\\ - {H}^{+} {{H}^{+}}^\dagger {\textcolor{blue}{\phi_{1}}}^{0} {{\textcolor{blue}{\phi_{1}}}^{0}}^\dagger + {H}^{+} {{H}^{+}}^\dagger {\textcolor{blue}{\phi_{1}}}^{++} {{\textcolor{blue}{\phi_{1}}}^{++}}^\dagger \\\\ + \sqrt{2}\, {H}^{+} {{H}^{0}}^\dagger {\textcolor{blue}{\phi_{1}}}^{0} {{\textcolor{blue}{\phi_{1}}}^{+}}^\dagger + \sqrt{2}\, {H}^{+} {{H}^{0}}^\dagger {\textcolor{blue}{\phi_{1}}}^{+} {{\textcolor{blue}{\phi_{1}}}^{++}}^\dagger\end{array} \right)
\end{dmath*}

\begin{dmath*}
\lambda_{5,c2}^{(2)}\left( \begin{array}{l}- \sqrt{2}\, {H}^{0} {{H}^{+}}^\dagger {\textcolor{blue}{\phi_{1}}}^{+} {{\textcolor{blue}{\phi_{1}}}^{0}}^\dagger - \sqrt{2}\, {H}^{0} {{H}^{+}}^\dagger {\textcolor{blue}{\phi_{1}}}^{++} {{\textcolor{blue}{\phi_{1}}}^{+}}^\dagger \\\\ + {H}^{0} {{H}^{0}}^\dagger {\textcolor{blue}{\phi_{1}}}^{+} {{\textcolor{blue}{\phi_{1}}}^{+}}^\dagger + 2\, {H}^{0} {{H}^{0}}^\dagger {\textcolor{blue}{\phi_{1}}}^{++} {{\textcolor{blue}{\phi_{1}}}^{++}}^\dagger \\\\ + 2\, {H}^{+} {{H}^{+}}^\dagger {\textcolor{blue}{\phi_{1}}}^{0} {{\textcolor{blue}{\phi_{1}}}^{0}}^\dagger + {H}^{+} {{H}^{+}}^\dagger {\textcolor{blue}{\phi_{1}}}^{+} {{\textcolor{blue}{\phi_{1}}}^{+}}^\dagger \\\\ - \sqrt{2}\, {H}^{+} {{H}^{0}}^\dagger {\textcolor{blue}{\phi_{1}}}^{0} {{\textcolor{blue}{\phi_{1}}}^{+}}^\dagger - \sqrt{2}\, {H}^{+} {{H}^{0}}^\dagger {\textcolor{blue}{\phi_{1}}}^{+} {{\textcolor{blue}{\phi_{1}}}^{++}}^\dagger\end{array} \right)
\end{dmath*}

\begin{dmath*}
\lambda_{7,c1}^{(1)}\left( \begin{array}{l}{\textcolor{blue}{\phi_{1}}}^{0} {\textcolor{blue}{\phi_{1}}}^{0} {{\textcolor{blue}{\phi_{1}}}^{0}}^\dagger {{\textcolor{blue}{\phi_{1}}}^{0}}^\dagger + 2\, {\textcolor{blue}{\phi_{1}}}^{0} {\textcolor{blue}{\phi_{1}}}^{+} {{\textcolor{blue}{\phi_{1}}}^{+}}^\dagger {{\textcolor{blue}{\phi_{1}}}^{0}}^\dagger \\\\ + 2\, {\textcolor{blue}{\phi_{1}}}^{0} {\textcolor{blue}{\phi_{1}}}^{++} {{\textcolor{blue}{\phi_{1}}}^{++}}^\dagger {{\textcolor{blue}{\phi_{1}}}^{0}}^\dagger + {\textcolor{blue}{\phi_{1}}}^{+} {\textcolor{blue}{\phi_{1}}}^{+} {{\textcolor{blue}{\phi_{1}}}^{+}}^\dagger {{\textcolor{blue}{\phi_{1}}}^{+}}^\dagger \\\\ + 2\, {\textcolor{blue}{\phi_{1}}}^{+} {\textcolor{blue}{\phi_{1}}}^{++} {{\textcolor{blue}{\phi_{1}}}^{++}}^\dagger {{\textcolor{blue}{\phi_{1}}}^{+}}^\dagger + {\textcolor{blue}{\phi_{1}}}^{++} {\textcolor{blue}{\phi_{1}}}^{++} {{\textcolor{blue}{\phi_{1}}}^{++}}^\dagger {{\textcolor{blue}{\phi_{1}}}^{++}}^\dagger\end{array} \right)
\end{dmath*}

\begin{dmath*}
\lambda_{7,c2}^{(2)}\left( \begin{array}{l}{\textcolor{blue}{\phi_{1}}}^{0} {\textcolor{blue}{\phi_{1}}}^{0} {{\textcolor{blue}{\phi_{1}}}^{0}}^\dagger {{\textcolor{blue}{\phi_{1}}}^{0}}^\dagger + 2\, {\textcolor{blue}{\phi_{1}}}^{0} {\textcolor{blue}{\phi_{1}}}^{+} {{\textcolor{blue}{\phi_{1}}}^{+}}^\dagger {{\textcolor{blue}{\phi_{1}}}^{0}}^\dagger \\\\ + {\textcolor{blue}{\phi_{1}}}^{0} {\textcolor{blue}{\phi_{1}}}^{++} {{\textcolor{blue}{\phi_{1}}}^{+}}^\dagger {{\textcolor{blue}{\phi_{1}}}^{+}}^\dagger + {\textcolor{blue}{\phi_{1}}}^{+} {\textcolor{blue}{\phi_{1}}}^{+} {{\textcolor{blue}{\phi_{1}}}^{++}}^\dagger {{\textcolor{blue}{\phi_{1}}}^{0}}^\dagger \\\\ + \frac{1}{2}\, {\textcolor{blue}{\phi_{1}}}^{+} {\textcolor{blue}{\phi_{1}}}^{+} {{\textcolor{blue}{\phi_{1}}}^{+}}^\dagger {{\textcolor{blue}{\phi_{1}}}^{+}}^\dagger + 2\, {\textcolor{blue}{\phi_{1}}}^{+} {\textcolor{blue}{\phi_{1}}}^{++} {{\textcolor{blue}{\phi_{1}}}^{++}}^\dagger {{\textcolor{blue}{\phi_{1}}}^{+}}^\dagger \\\\ + {\textcolor{blue}{\phi_{1}}}^{++} {\textcolor{blue}{\phi_{1}}}^{++} {{\textcolor{blue}{\phi_{1}}}^{++}}^\dagger {{\textcolor{blue}{\phi_{1}}}^{++}}^\dagger\end{array} \right)
\end{dmath*}
\end{pdf}

Now suppose that a user has independently constructed the same model and wishes to map their chosen basis onto the basis used by the package. To illustrate this, let us assume that the user has defined the following field as a two-index symmetric tensor:
\begin{align*}
    \Phi(1,3,1)= \Phi^{ab} = \Phi^{ba},
\end{align*}
where $a,b= 1,2$ are the $SU(2)$ indices. Moreover, suppose that the user has chosen the following phase conventions and normalizations for the component fields:  
\begin{align*}
\Phi^{11} = -\Phi^{++}, \quad   \Phi^{12} =\frac{1}{\sqrt{2}} \Phi^{+}, \quad   \Phi^{22} = \Phi^{0}.
\end{align*}
Then it is trivial to see that the field is canonically normalized
\begin{align*}
\Phi^{ab} \Phi^*_{ab} = \Phi^{++}\Phi^{++} + \Phi^{+}\Phi^{-} + \Phi^{0}\Phi^{0\ast}.
\end{align*}

The Yukawa interaction (by suppressing the Lorentz indices):
\begin{align*}
Y_S^{ij} L^{a}_i L^{b}_j \epsilon_{ac} \epsilon_{bd} \Phi^{cd}  =-Y_S^{ij} \Phi^{++} e_i e_j  -Y_S^{ij} \frac{1}{\sqrt{2}} \Phi^{-} \left( e_i \nu_j+ \nu_i e_j  \right) +Y_S^{ij} \Phi^{0} \nu_i \nu_j.
\end{align*}

Furthermore, the new physics interactions in the scalar potential are given by
\begin{align*}
V&\supset \omega_1 (H^\ast_a H^a)(\Phi^\ast_{bc} \Phi^{bc}) + \omega_2  H^\ast_a H^b \Phi^\ast_{bc} \Phi^{ca} +\bigg\{ \zeta_1  H^aH^b \Phi^\ast_{ab} +\mathrm{h.c.} \bigg\} 
\\
&+ 
\Lambda_1 (\Phi^\ast_{bc} \Phi^{bc})^2 + \Lambda_2 \Phi^\ast_{ab} \Phi^{bc} \Phi^\ast_{cd} \Phi^{da}.
\end{align*}
Since the user's explicit conventions have been specified, the reader can straightforwardly expand the above Lagrangian. By comparing the resulting expressions with those produced by the package, one can verify that all Lagrangian terms agree, provided that the following mapping is applied. In particular, the required phase redefinition is:
\begin{align*}
&\Phi^{\pm \pm}\to - {{\textcolor{blue}{\phi_{1}}}^{\pm \pm}}
\end{align*}
as well as with the following mapping:
\begin{align*}
& Y_3 = Y_S
\\
& \mu_3 = \zeta_1,
\\
&\lambda_{5,c1} = 2 \omega_2 - \omega_1
\\
&\lambda_{5,c2} = 2 \omega_1 - 2\omega_2 
\\
&\lambda_{7,c1} = \Lambda_1
\\
&\lambda_{7,c2} = \Lambda_2 
\end{align*}

In summary, the expansion of Lagrangian terms can only be performed after specifying a particular choice of basis, phase conventions, and field normalizations. Consequently, direct comparisons between expanded expressions are meaningful only after accounting for these conventions. The user can always construct an appropriate mapping between their conventions and those employed by the package.

\section{A model of Leptoquarks}\label{sec:08}

Let us now construct a more involved model to illustrate the power and versatility of the package introduced in this work. This model consists of all scalar leptoquarks. As the generated report will show, the model contains
\begin{itemize}
\item Kinetic terms: 14
\item Scalar potential terms: 73
\item Yukawa interaction terms: 14
\end{itemize}
for a total of more than 100 terms in the Lagrangian.

Constructing such a model manually is a lengthy and laborious process that is prone to human error. Furthermore, expanding and verifying such a large number of terms can be both time-consuming and cumbersome. In contrast, the package is able to construct the complete Lagrangian, identify all independent interactions, and perform the corresponding expansions in less than a minute.

\subsection{Model construction}
We begin constructing the model by adding a scalar leptoquark with the quantum numbers $(3,1,-1/3)$: 
\begin{user}
    build a model with a scalar field, which has hypercharge minus one-third, weak singlet, and color triplet
\end{user}

\begin{agent}
    I interpreted the BSM field content as scalar (3,1,-1/3). If this is correct, type only yes. If not, restate the quantum numbers.
\end{agent}

\begin{user}
    yes
\end{user}

\begin{agent}
    The model is constructed. [0.1s]
\end{agent}

Second, we add to the model a scalar leptoquark $(3,1,-4/3)$: 
\begin{user}
    add a scalar, color triplet, weak singlet, hypercharge minus four third
\end{user}

\begin{agent}
    I interpreted the BSM field content as scalar (3,1,-4/3). If this is correct, type only yes. If not, restate the quantum numbers.
\end{agent}

\begin{user}
    yes
\end{user}

\begin{agent}
    The model is constructed. [0.1s]
\end{agent}

Next, we add a scalar leptoquark $(3,2,7/6)$:
\begin{user}
    add another scalar (3,2,7/6)
\end{user}

\begin{agent}
    I interpreted the BSM field content as scalar (3,2,7/6). If this is correct, type only yes. If not, restate the quantum numbers.
\end{agent}

\begin{user}
    yes
\end{user}

\begin{agent}
    The model is constructed. [0.1s]
\end{agent}

Furthermore, we introduce the scalar leptoquark $(3,2,1/6)$
\begin{user}
    introduce yet another field, scalar, color triplet, weak doublet, hypercharge 1/6
\end{user}

\begin{agent}
    I interpreted the BSM field content as scalar (3,2,1/6). If this is correct, type only yes. If not, restate the quantum numbers.
\end{agent}

\begin{agent}
    The model is constructed. [0.1s]
\end{agent}

The last scalar leptoquark we insert to the model is $(3,3,-1/3)$:
\begin{user}
    finally, add a scalar with color triplet, weak triplet, hypercharge= -1/3
\end{user}

\begin{agent}
    I interpreted the BSM field content as scalar (3,3,-1/3). If this is correct, type only yes. If not, restate the quantum numbers.
\end{agent}

\begin{user}
    yes
\end{user}

\begin{agent}
    The model is constructed. [1.7s]
\end{agent}

\subsection{Generating PDF}
\begin{user}
    generate pdf
\end{user}

\begin{agent}
\begin{verbatim}
Generated report for 'SM+phi_1+phi_2+phi_3+phi_4+phi_5'
with field content (3,1,-1/3), (3,1,-4/3), (3,2,7/6), (3,2,1/6), (3,3,-1/3). 
TeX: /home/user/Downloads/bsm_agent/output/SM_phi_1_phi_2_phi_3_phi_4_phi_5.tex. 
PDF: /home/user/Downloads/bsm_agent/output/SM_phi_1_phi_2_phi_3_phi_4_phi_5.pdf. 
[1.4s]
\end{verbatim}
\end{agent}

As before, when it is asked to produce a PDF, it creates two sections:
\begin{pdf}
\centering
\includegraphics[width=0.6\textwidth]{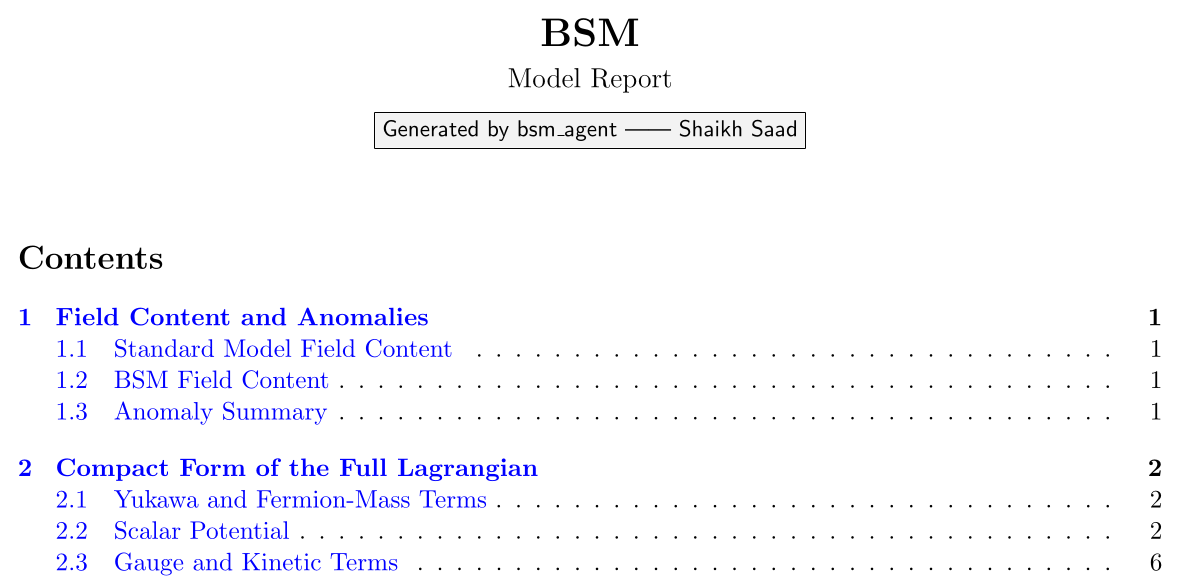}
\end{pdf}
The first section reads as:
\begin{pdf}
BSM Field Content:\
\[
\begin{aligned}
\textcolor{blue}{\phi_{1}} & = & \text{scalar} & \quad (3,1,-1/3) \\
\textcolor{blue}{\phi_{2}} & = & \text{scalar} & \quad (3,1,-4/3) \\
\textcolor{blue}{\phi_{3}} & = & \text{scalar} & \quad (3,2,7/6) \\
\textcolor{blue}{\phi_{4}} & = & \text{scalar} & \quad (3,2,1/6) \\
\textcolor{blue}{\phi_{5}} & = & \text{scalar} & \quad (3,3,-1/3)
\end{aligned}
\]

Anomaly Summary:\\
\textbf{Anomalies:} \textcolor{green!50!black}{\textbf{free}}.
\end{pdf}
Whereas, the second section writes all the Lagrangian terms in compact form.

The complete Yukawa sector is given by:
\begin{pdf}
\begingroup
\allowdisplaybreaks
\begin{align*}
\mathcal{L}_{Y} = & Y_{1} \, q q \textcolor{blue}{\phi_{1}} + Y_{2} \, q q \textcolor{blue}{\phi_{5}} + Y_{3} \, \textcolor{blue}{\phi_{1}}^\dagger l q + Y_{4} \, \textcolor{blue}{\phi_{5}}^\dagger l q \\
& + Y_{5} \, H^\dagger d^C q + Y_{6} \, u^C q H + Y_{7} \, \textcolor{blue}{\phi_{3}}^\dagger e^C q + Y_{8} \, d^C l \textcolor{blue}{\phi_{4}} \\
& + Y_{9} \, u^C l \textcolor{blue}{\phi_{3}} + Y_{10} \, H^\dagger e^C l + Y_{11} \, \textcolor{blue}{\phi_{1}}^\dagger d^C u^C + Y_{12} \, e^C d^C \textcolor{blue}{\phi_{2}} \\
& + Y_{13} \, \textcolor{blue}{\phi_{2}}^\dagger u^C u^C + Y_{14} \, e^C u^C \textcolor{blue}{\phi_{1}} + \mathrm{h.c.}
\end{align*}
\endgroup
\end{pdf}
which contains 14 different terms. While, the scalar potential composed of 73 terms:
\begin{pdf}
\begingroup
\allowdisplaybreaks
\begin{align*}
V = & m^{2}_{1} \, H H^\dagger + m^{2}_{2} \, \textcolor{blue}{\phi_{1}} \textcolor{blue}{\phi_{1}}^\dagger + m^{2}_{3} \, \textcolor{blue}{\phi_{2}} \textcolor{blue}{\phi_{2}}^\dagger + m^{2}_{4} \, \textcolor{blue}{\phi_{3}} \textcolor{blue}{\phi_{3}}^\dagger \\
& + m^{2}_{5} \, \textcolor{blue}{\phi_{4}} \textcolor{blue}{\phi_{4}}^\dagger + m^{2}_{6} \, \textcolor{blue}{\phi_{5}} \textcolor{blue}{\phi_{5}}^\dagger + \bigg\{ \mu_{7} \, H \textcolor{blue}{\phi_{1}} \textcolor{blue}{\phi_{4}}^\dagger + \mu_{8} \, H \textcolor{blue}{\phi_{4}}^\dagger \textcolor{blue}{\phi_{5}} \\
& + \mu_{9} \, \textcolor{blue}{\phi_{1}} \textcolor{blue}{\phi_{4}} \textcolor{blue}{\phi_{4}} + \mu_{10} \, \textcolor{blue}{\phi_{2}} \textcolor{blue}{\phi_{3}} \textcolor{blue}{\phi_{4}} + \mathrm{h.c.} \bigg\}  + \lambda_{11} \, H H H^\dagger H^\dagger +\bigg\{ \lambda_{12} \, H H \textcolor{blue}{\phi_{2}} \textcolor{blue}{\phi_{5}}^\dagger \\
& + \lambda_{13} \, H H \textcolor{blue}{\phi_{3}}^\dagger \textcolor{blue}{\phi_{4}} + \mathrm{h.c.} \bigg\} + \lambda_{14} \, H H^\dagger \textcolor{blue}{\phi_{1}} \textcolor{blue}{\phi_{1}}^\dagger + \bigg\{ \lambda_{15} \, H H^\dagger \textcolor{blue}{\phi_{1}} \textcolor{blue}{\phi_{5}}^\dagger + \mathrm{h.c.} \bigg\} + \lambda_{16} \, H H^\dagger \textcolor{blue}{\phi_{2}} \textcolor{blue}{\phi_{2}}^\dagger \\
& + \lambda_{17,c1} \, H H^\dagger \textcolor{blue}{\phi_{3}} \textcolor{blue}{\phi_{3}}^\dagger + \lambda_{17,c2} \, H H^\dagger \textcolor{blue}{\phi_{3}} \textcolor{blue}{\phi_{3}}^\dagger + \lambda_{19,c1} \, H H^\dagger \textcolor{blue}{\phi_{4}} \textcolor{blue}{\phi_{4}}^\dagger + \lambda_{19,c2} \, H H^\dagger \textcolor{blue}{\phi_{4}} \textcolor{blue}{\phi_{4}}^\dagger \\
& + \lambda_{21,c1} \, H H^\dagger \textcolor{blue}{\phi_{5}} \textcolor{blue}{\phi_{5}}^\dagger + \lambda_{21,c2} \, H H^\dagger \textcolor{blue}{\phi_{5}} \textcolor{blue}{\phi_{5}}^\dagger + \bigg\{ \lambda_{23} \, H \textcolor{blue}{\phi_{1}} \textcolor{blue}{\phi_{2}} \textcolor{blue}{\phi_{3}} + \lambda_{24} \, H \textcolor{blue}{\phi_{1}} \textcolor{blue}{\phi_{4}} \textcolor{blue}{\phi_{5}} \\
& + \lambda_{25} \, H \textcolor{blue}{\phi_{1}}^\dagger \textcolor{blue}{\phi_{3}}^\dagger \textcolor{blue}{\phi_{5}}^\dagger + \lambda_{26} \, H \textcolor{blue}{\phi_{2}} \textcolor{blue}{\phi_{3}} \textcolor{blue}{\phi_{5}} + \lambda_{27} \, H \textcolor{blue}{\phi_{3}}^\dagger \textcolor{blue}{\phi_{5}}^\dagger \textcolor{blue}{\phi_{5}}^\dagger + \lambda_{28} \, H \textcolor{blue}{\phi_{4}} \textcolor{blue}{\phi_{5}} \textcolor{blue}{\phi_{5}}  + \mathrm{h.c.} \bigg\} \\
& + \lambda_{29} \, \textcolor{blue}{\phi_{1}} \textcolor{blue}{\phi_{1}} \textcolor{blue}{\phi_{1}}^\dagger \textcolor{blue}{\phi_{1}}^\dagger + \bigg\{ \lambda_{30} \, \textcolor{blue}{\phi_{1}} \textcolor{blue}{\phi_{1}} \textcolor{blue}{\phi_{5}}^\dagger \textcolor{blue}{\phi_{5}}^\dagger + \mathrm{h.c.} \bigg\} + \lambda_{31,c1} \, \textcolor{blue}{\phi_{1}} \textcolor{blue}{\phi_{1}}^\dagger \textcolor{blue}{\phi_{2}} \textcolor{blue}{\phi_{2}}^\dagger + \lambda_{31,c2} \, \textcolor{blue}{\phi_{1}} \textcolor{blue}{\phi_{1}}^\dagger \textcolor{blue}{\phi_{2}} \textcolor{blue}{\phi_{2}}^\dagger \\
& + \lambda_{33,c1} \, \textcolor{blue}{\phi_{1}} \textcolor{blue}{\phi_{1}}^\dagger \textcolor{blue}{\phi_{3}} \textcolor{blue}{\phi_{3}}^\dagger + \lambda_{33,c2} \, \textcolor{blue}{\phi_{1}} \textcolor{blue}{\phi_{1}}^\dagger \textcolor{blue}{\phi_{3}} \textcolor{blue}{\phi_{3}}^\dagger + \lambda_{35,c1} \, \textcolor{blue}{\phi_{1}} \textcolor{blue}{\phi_{1}}^\dagger \textcolor{blue}{\phi_{4}} \textcolor{blue}{\phi_{4}}^\dagger + \lambda_{35,c2} \, \textcolor{blue}{\phi_{1}} \textcolor{blue}{\phi_{1}}^\dagger \textcolor{blue}{\phi_{4}} \textcolor{blue}{\phi_{4}}^\dagger \\
& + \lambda_{37,c1} \, \textcolor{blue}{\phi_{1}} \textcolor{blue}{\phi_{1}}^\dagger \textcolor{blue}{\phi_{5}} \textcolor{blue}{\phi_{5}}^\dagger + \lambda_{37,c2} \, \textcolor{blue}{\phi_{1}} \textcolor{blue}{\phi_{1}}^\dagger \textcolor{blue}{\phi_{5}} \textcolor{blue}{\phi_{5}}^\dagger + \bigg\{ \lambda_{39,c1} \, \textcolor{blue}{\phi_{1}} \textcolor{blue}{\phi_{2}}^\dagger \textcolor{blue}{\phi_{3}}^\dagger \textcolor{blue}{\phi_{4}} + \lambda_{39,c2} \, \textcolor{blue}{\phi_{1}} \textcolor{blue}{\phi_{2}}^\dagger \textcolor{blue}{\phi_{3}}^\dagger \textcolor{blue}{\phi_{4}} \\
& + \lambda_{41,c1} \, \textcolor{blue}{\phi_{1}} \textcolor{blue}{\phi_{3}} \textcolor{blue}{\phi_{3}}^\dagger \textcolor{blue}{\phi_{5}}^\dagger + \lambda_{41,c2} \, \textcolor{blue}{\phi_{1}} \textcolor{blue}{\phi_{3}} \textcolor{blue}{\phi_{3}}^\dagger \textcolor{blue}{\phi_{5}}^\dagger + \lambda_{43,c1} \, \textcolor{blue}{\phi_{1}} \textcolor{blue}{\phi_{4}} \textcolor{blue}{\phi_{4}}^\dagger \textcolor{blue}{\phi_{5}}^\dagger + \lambda_{43,c2} \, \textcolor{blue}{\phi_{1}} \textcolor{blue}{\phi_{4}} \textcolor{blue}{\phi_{4}}^\dagger \textcolor{blue}{\phi_{5}}^\dagger \\
& + \lambda_{45} \, \textcolor{blue}{\phi_{1}} \textcolor{blue}{\phi_{5}} \textcolor{blue}{\phi_{5}}^\dagger \textcolor{blue}{\phi_{5}}^\dagger + \mathrm{h.c.} \bigg\} + \lambda_{46} \, \textcolor{blue}{\phi_{2}} \textcolor{blue}{\phi_{2}} \textcolor{blue}{\phi_{2}}^\dagger \textcolor{blue}{\phi_{2}}^\dagger + \lambda_{47,c1} \, \textcolor{blue}{\phi_{2}} \textcolor{blue}{\phi_{2}}^\dagger \textcolor{blue}{\phi_{3}} \textcolor{blue}{\phi_{3}}^\dagger + \lambda_{47,c2} \, \textcolor{blue}{\phi_{2}} \textcolor{blue}{\phi_{2}}^\dagger \textcolor{blue}{\phi_{3}} \textcolor{blue}{\phi_{3}}^\dagger \\
& + \lambda_{49,c1} \, \textcolor{blue}{\phi_{2}} \textcolor{blue}{\phi_{2}}^\dagger \textcolor{blue}{\phi_{4}} \textcolor{blue}{\phi_{4}}^\dagger + \lambda_{49,c2} \, \textcolor{blue}{\phi_{2}} \textcolor{blue}{\phi_{2}}^\dagger \textcolor{blue}{\phi_{4}} \textcolor{blue}{\phi_{4}}^\dagger + \lambda_{51,c1} \, \textcolor{blue}{\phi_{2}} \textcolor{blue}{\phi_{2}}^\dagger \textcolor{blue}{\phi_{5}} \textcolor{blue}{\phi_{5}}^\dagger + \lambda_{51,c2} \, \textcolor{blue}{\phi_{2}} \textcolor{blue}{\phi_{2}}^\dagger \textcolor{blue}{\phi_{5}} \textcolor{blue}{\phi_{5}}^\dagger \\
& + \bigg\{ \lambda_{53,c1} \, \textcolor{blue}{\phi_{2}} \textcolor{blue}{\phi_{3}} \textcolor{blue}{\phi_{4}}^\dagger \textcolor{blue}{\phi_{5}}^\dagger + \lambda_{53,c2} \, \textcolor{blue}{\phi_{2}} \textcolor{blue}{\phi_{3}} \textcolor{blue}{\phi_{4}}^\dagger \textcolor{blue}{\phi_{5}}^\dagger + \mathrm{h.c.} \bigg\} + \lambda_{55,c1} \, \textcolor{blue}{\phi_{3}} \textcolor{blue}{\phi_{3}} \textcolor{blue}{\phi_{3}}^\dagger \textcolor{blue}{\phi_{3}}^\dagger + \lambda_{55,c2} \, \textcolor{blue}{\phi_{3}} \textcolor{blue}{\phi_{3}} \textcolor{blue}{\phi_{3}}^\dagger \textcolor{blue}{\phi_{3}}^\dagger \\
& + \lambda_{57,c1} \, \textcolor{blue}{\phi_{3}} \textcolor{blue}{\phi_{3}}^\dagger \textcolor{blue}{\phi_{4}} \textcolor{blue}{\phi_{4}}^\dagger + \lambda_{57,c2} \, \textcolor{blue}{\phi_{3}} \textcolor{blue}{\phi_{3}}^\dagger \textcolor{blue}{\phi_{4}} \textcolor{blue}{\phi_{4}}^\dagger + \lambda_{57,c3} \, \textcolor{blue}{\phi_{3}} \textcolor{blue}{\phi_{3}}^\dagger \textcolor{blue}{\phi_{4}} \textcolor{blue}{\phi_{4}}^\dagger + \lambda_{57,c4} \, \textcolor{blue}{\phi_{3}} \textcolor{blue}{\phi_{3}}^\dagger \textcolor{blue}{\phi_{4}} \textcolor{blue}{\phi_{4}}^\dagger \\
& + \lambda_{61,c1} \, \textcolor{blue}{\phi_{3}} \textcolor{blue}{\phi_{3}}^\dagger \textcolor{blue}{\phi_{5}} \textcolor{blue}{\phi_{5}}^\dagger + \lambda_{61,c2} \, \textcolor{blue}{\phi_{3}} \textcolor{blue}{\phi_{3}}^\dagger \textcolor{blue}{\phi_{5}} \textcolor{blue}{\phi_{5}}^\dagger + \lambda_{61,c3} \, \textcolor{blue}{\phi_{3}} \textcolor{blue}{\phi_{3}}^\dagger \textcolor{blue}{\phi_{5}} \textcolor{blue}{\phi_{5}}^\dagger + \lambda_{61,c4} \, \textcolor{blue}{\phi_{3}} \textcolor{blue}{\phi_{3}}^\dagger \textcolor{blue}{\phi_{5}} \textcolor{blue}{\phi_{5}}^\dagger \\
& + \lambda_{65,c1} \, \textcolor{blue}{\phi_{4}} \textcolor{blue}{\phi_{4}} \textcolor{blue}{\phi_{4}}^\dagger \textcolor{blue}{\phi_{4}}^\dagger + \lambda_{65,c2} \, \textcolor{blue}{\phi_{4}} \textcolor{blue}{\phi_{4}} \textcolor{blue}{\phi_{4}}^\dagger \textcolor{blue}{\phi_{4}}^\dagger + \lambda_{67,c1} \, \textcolor{blue}{\phi_{4}} \textcolor{blue}{\phi_{4}}^\dagger \textcolor{blue}{\phi_{5}} \textcolor{blue}{\phi_{5}}^\dagger + \lambda_{67,c2} \, \textcolor{blue}{\phi_{4}} \textcolor{blue}{\phi_{4}}^\dagger \textcolor{blue}{\phi_{5}} \textcolor{blue}{\phi_{5}}^\dagger \\
& + \lambda_{67,c3} \, \textcolor{blue}{\phi_{4}} \textcolor{blue}{\phi_{4}}^\dagger \textcolor{blue}{\phi_{5}} \textcolor{blue}{\phi_{5}}^\dagger + \lambda_{67,c4} \, \textcolor{blue}{\phi_{4}} \textcolor{blue}{\phi_{4}}^\dagger \textcolor{blue}{\phi_{5}} \textcolor{blue}{\phi_{5}}^\dagger + \lambda_{71,c1} \, \textcolor{blue}{\phi_{5}} \textcolor{blue}{\phi_{5}} \textcolor{blue}{\phi_{5}}^\dagger \textcolor{blue}{\phi_{5}}^\dagger + \lambda_{71,c2} \, \textcolor{blue}{\phi_{5}} \textcolor{blue}{\phi_{5}} \textcolor{blue}{\phi_{5}}^\dagger \textcolor{blue}{\phi_{5}}^\dagger \\
& + \lambda_{71,c3} \, \textcolor{blue}{\phi_{5}} \textcolor{blue}{\phi_{5}} \textcolor{blue}{\phi_{5}}^\dagger \textcolor{blue}{\phi_{5}}^\dagger  
\end{align*}
\endgroup
\end{pdf}
And the 14 gauge kinetic terms:
\begin{pdf}
\begingroup
\allowdisplaybreaks
\begin{align*}
\mathcal{L}_{\text{kin}} = & -\frac{1}{4} B_{\mu\nu} B^{\mu\nu} -\frac{1}{4} W^a_{\mu\nu} W^{a,\mu\nu} -\frac{1}{4} G^A_{\mu\nu} G^{A,\mu\nu} + (D_\mu q)^\dagger D^\mu q \\
& + (D_\mu l)^\dagger D^\mu l + (D_\mu d^C)^\dagger D^\mu d^C + (D_\mu u^C)^\dagger D^\mu u^C + (D_\mu e^C)^\dagger D^\mu e^C \\
& + (D_\mu H)^\dagger D^\mu H + (D_\mu \textcolor{blue}{\phi_{1}})^\dagger D^\mu \textcolor{blue}{\phi_{1}} + (D_\mu \textcolor{blue}{\phi_{2}})^\dagger D^\mu \textcolor{blue}{\phi_{2}} + (D_\mu \textcolor{blue}{\phi_{3}})^\dagger D^\mu \textcolor{blue}{\phi_{3}} \\
& + (D_\mu \textcolor{blue}{\phi_{4}})^\dagger D^\mu \textcolor{blue}{\phi_{4}} + (D_\mu \textcolor{blue}{\phi_{5}})^\dagger D^\mu \textcolor{blue}{\phi_{5}}
\end{align*}
\endgroup
\end{pdf}

\subsection{Expanding the Full Lagrangian}

Once the model has been constructed, the next important step is to analyze its phenomenological content by expanding the Lagrangian terms. The user can request the package to perform this expansion:
\begin{user}
    expand the Lagrangian
\end{user}

\begin{agent}
\begin{verbatim}
Generated report for 'SM+phi_1+phi_2+phi_3+phi_4+phi_5'
with field content (3,1,-1/3), (3,1,-4/3), (3,2,7/6), (3,2,1/6), (3,3,-1/3). 
TeX: /home/user/Downloads/bsm_agent/output/SM_phi_1_phi_2_phi_3_phi_4_phi_5.tex. 
PDF: /home/user/Downloads/bsm_agent/output/SM_phi_1_phi_2_phi_3_phi_4_phi_5.pdf. 
[7.8s]
\end{verbatim}
\end{agent}
As a result, the package has expanded all the Lagrangian terms in the $SU(3)$ and $SU(2)$ component forms.
\begin{pdf}
\centering
\includegraphics[width=0.65\textwidth]{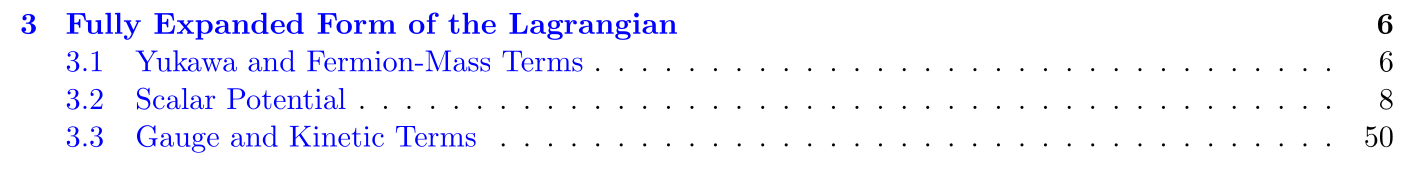}
\end{pdf}
Owing to the large number of terms in the model, we do not display all expanded expressions in this manuscript. Instead, a PDF containing the full expansion of the Lagrangian is provided in the GitHub repository: \href{https://github.com/saad-hep/bsm_agent}{\faGithub~bsm_agent}.

\subsubsection{Expanded Yukawa Sector}
Let us examine a few representative terms in detail. For example, consider the Yukawa interactions involving the leptoquarks (LQs) $\textcolor{blue}{\phi_{3}}$ and $\textcolor{blue}{\phi_{5}}$:
\begin{pdf}
\begin{align*}
\mathcal{L}_{Y} \supset  Y_{7} \, \textcolor{blue}{\phi_{3}}^\dagger e^C q  + Y_{9} \, u^C l \textcolor{blue}{\phi_{3}} + \mathrm{h.c.},
\end{align*}    
\end{pdf}
and
\begin{pdf}
\begin{align*}
\mathcal{L}_{Y} \supset Y_{2} \, q q \textcolor{blue}{\phi_{5}} + Y_{4} \, \textcolor{blue}{\phi_{5}}^\dagger l q + \mathrm{h.c.},
\end{align*}    
\end{pdf}
respectively. For $\textcolor{blue}{\phi_{3}}(3,2,7/6)$ LQ, the package reports the expanded terms as:
\begin{pdf}
\begingroup
\allowdisplaybreaks    
\begin{dmath*}
Y_{7}\left( \begin{array}{l}{{\textcolor{blue}{\phi_{3}}}^{+2/3,\bar{b}}}^\dagger {e^C}^{+} {q}^{-1/3}_{b} + {{\textcolor{blue}{\phi_{3}}}^{+2/3,\bar{g}}}^\dagger {e^C}^{+} {q}^{-1/3}_{g} \\\\ + {{\textcolor{blue}{\phi_{3}}}^{+2/3,\bar{r}}}^\dagger {e^C}^{+} {q}^{-1/3}_{r} + {{\textcolor{blue}{\phi_{3}}}^{+5/3,\bar{b}}}^\dagger {e^C}^{+} {q}^{+2/3}_{b} \\\\ + {{\textcolor{blue}{\phi_{3}}}^{+5/3,\bar{g}}}^\dagger {e^C}^{+} {q}^{+2/3}_{g} + {{\textcolor{blue}{\phi_{3}}}^{+5/3,\bar{r}}}^\dagger {e^C}^{+} {q}^{+2/3}_{r}\end{array} \right) + \mathrm{h.c.}
\end{dmath*}
\begin{dmath*}
Y_{9}\left( \begin{array}{l}- {u^C}^{-2/3,\bar{b}} {l}^{-} {\textcolor{blue}{\phi_{3}}}^{+5/3}_{b} + {u^C}^{-2/3,\bar{b}} {l}^{0} {\textcolor{blue}{\phi_{3}}}^{+2/3}_{b} \\\\ - {u^C}^{-2/3,\bar{g}} {l}^{-} {\textcolor{blue}{\phi_{3}}}^{+5/3}_{g} + {u^C}^{-2/3,\bar{g}} {l}^{0} {\textcolor{blue}{\phi_{3}}}^{+2/3}_{g} \\\\ - {u^C}^{-2/3,\bar{r}} {l}^{-} {\textcolor{blue}{\phi_{3}}}^{+5/3}_{r} + {u^C}^{-2/3,\bar{r}} {l}^{0} {\textcolor{blue}{\phi_{3}}}^{+2/3}_{r}\end{array} \right) + \mathrm{h.c.}
\end{dmath*}
\endgroup
\end{pdf}
where, for $Y_9$, the relative minus sign between the $u^C e^-$ and $u^C\nu$ is apparent; see, for example, Eq.(4) of Ref.~\cite{Dorsner:2016wpm}. Here, Yukawa couplings, $Y_{7,9}$ are arbitrary $3\times 3$ matrices in the flavor space.

As for the $\textcolor{blue}{\phi_{5}}(3,3,-1/3)$ LQ, the package reports:
\begin{pdf}
\begingroup
\allowdisplaybreaks 
\begin{dmath*}
Y_{2}\left( \begin{array}{l}- {q}^{+2/3}_{b} {q}^{+2/3}_{g} {\textcolor{blue}{\phi_{5}}}^{-4/3}_{r} + {q}^{+2/3}_{b} {q}^{+2/3}_{r} {\textcolor{blue}{\phi_{5}}}^{-4/3}_{g} \\\\ + \frac{\sqrt{2}}{2}\, {q}^{+2/3}_{b} {q}^{-1/3}_{g} {\textcolor{blue}{\phi_{5}}}^{-1/3}_{r} - \frac{\sqrt{2}}{2}\, {q}^{+2/3}_{b} {q}^{-1/3}_{r} {\textcolor{blue}{\phi_{5}}}^{-1/3}_{g} \\\\ + {q}^{+2/3}_{g} {q}^{+2/3}_{b} {\textcolor{blue}{\phi_{5}}}^{-4/3}_{r} - {q}^{+2/3}_{g} {q}^{+2/3}_{r} {\textcolor{blue}{\phi_{5}}}^{-4/3}_{b} \\\\ - \frac{\sqrt{2}}{2}\, {q}^{+2/3}_{g} {q}^{-1/3}_{b} {\textcolor{blue}{\phi_{5}}}^{-1/3}_{r} + \frac{\sqrt{2}}{2}\, {q}^{+2/3}_{g} {q}^{-1/3}_{r} {\textcolor{blue}{\phi_{5}}}^{-1/3}_{b} \\\\ - {q}^{+2/3}_{r} {q}^{+2/3}_{b} {\textcolor{blue}{\phi_{5}}}^{-4/3}_{g} + {q}^{+2/3}_{r} {q}^{+2/3}_{g} {\textcolor{blue}{\phi_{5}}}^{-4/3}_{b} \\\\ + \frac{\sqrt{2}}{2}\, {q}^{+2/3}_{r} {q}^{-1/3}_{b} {\textcolor{blue}{\phi_{5}}}^{-1/3}_{g} - \frac{\sqrt{2}}{2}\, {q}^{+2/3}_{r} {q}^{-1/3}_{g} {\textcolor{blue}{\phi_{5}}}^{-1/3}_{b} \\\\ + \frac{\sqrt{2}}{2}\, {q}^{-1/3}_{b} {q}^{+2/3}_{g} {\textcolor{blue}{\phi_{5}}}^{-1/3}_{r} - \frac{\sqrt{2}}{2}\, {q}^{-1/3}_{b} {q}^{+2/3}_{r} {\textcolor{blue}{\phi_{5}}}^{-1/3}_{g} \\\\ - {q}^{-1/3}_{b} {q}^{-1/3}_{g} {\textcolor{blue}{\phi_{5}}}^{+2/3}_{r} + {q}^{-1/3}_{b} {q}^{-1/3}_{r} {\textcolor{blue}{\phi_{5}}}^{+2/3}_{g} \\\\ - \frac{\sqrt{2}}{2}\, {q}^{-1/3}_{g} {q}^{+2/3}_{b} {\textcolor{blue}{\phi_{5}}}^{-1/3}_{r} + \frac{\sqrt{2}}{2}\, {q}^{-1/3}_{g} {q}^{+2/3}_{r} {\textcolor{blue}{\phi_{5}}}^{-1/3}_{b} \\\\ + {q}^{-1/3}_{g} {q}^{-1/3}_{b} {\textcolor{blue}{\phi_{5}}}^{+2/3}_{r} - {q}^{-1/3}_{g} {q}^{-1/3}_{r} {\textcolor{blue}{\phi_{5}}}^{+2/3}_{b} \\\\ + \frac{\sqrt{2}}{2}\, {q}^{-1/3}_{r} {q}^{+2/3}_{b} {\textcolor{blue}{\phi_{5}}}^{-1/3}_{g} - \frac{\sqrt{2}}{2}\, {q}^{-1/3}_{r} {q}^{+2/3}_{g} {\textcolor{blue}{\phi_{5}}}^{-1/3}_{b} \\\\ - {q}^{-1/3}_{r} {q}^{-1/3}_{b} {\textcolor{blue}{\phi_{5}}}^{+2/3}_{g} + {q}^{-1/3}_{r} {q}^{-1/3}_{g} {\textcolor{blue}{\phi_{5}}}^{+2/3}_{b}\end{array} \right) + \mathrm{h.c.}
\end{dmath*}
\begin{dmath*}
Y_{4}\left( \begin{array}{l}{{\textcolor{blue}{\phi_{5}}}^{+2/3,\bar{b}}}^\dagger {l}^{0} {q}^{+2/3}_{b} + {{\textcolor{blue}{\phi_{5}}}^{+2/3,\bar{g}}}^\dagger {l}^{0} {q}^{+2/3}_{g} \\\\ + {{\textcolor{blue}{\phi_{5}}}^{+2/3,\bar{r}}}^\dagger {l}^{0} {q}^{+2/3}_{r} + \frac{\sqrt{2}}{2}\, {{\textcolor{blue}{\phi_{5}}}^{-1/3,\bar{b}}}^\dagger {l}^{-} {q}^{+2/3}_{b} \\\\ + \frac{\sqrt{2}}{2}\, {{\textcolor{blue}{\phi_{5}}}^{-1/3,\bar{b}}}^\dagger {l}^{0} {q}^{-1/3}_{b} + \frac{\sqrt{2}}{2}\, {{\textcolor{blue}{\phi_{5}}}^{-1/3,\bar{g}}}^\dagger {l}^{-} {q}^{+2/3}_{g} \\\\ + \frac{\sqrt{2}}{2}\, {{\textcolor{blue}{\phi_{5}}}^{-1/3,\bar{g}}}^\dagger {l}^{0} {q}^{-1/3}_{g} + \frac{\sqrt{2}}{2}\, {{\textcolor{blue}{\phi_{5}}}^{-1/3,\bar{r}}}^\dagger {l}^{-} {q}^{+2/3}_{r} \\\\ + \frac{\sqrt{2}}{2}\, {{\textcolor{blue}{\phi_{5}}}^{-1/3,\bar{r}}}^\dagger {l}^{0} {q}^{-1/3}_{r} + {{\textcolor{blue}{\phi_{5}}}^{-4/3,\bar{b}}}^\dagger {l}^{-} {q}^{-1/3}_{b} \\\\ + {{\textcolor{blue}{\phi_{5}}}^{-4/3,\bar{g}}}^\dagger {l}^{-} {q}^{-1/3}_{g} + {{\textcolor{blue}{\phi_{5}}}^{-4/3,\bar{r}}}^\dagger {l}^{-} {q}^{-1/3}_{r}\end{array} \right) + \mathrm{h.c.}
\end{dmath*}
\endgroup
\end{pdf}

First, we scrutinize the $Y_2$ term (it is enough to consider only the $r$-colored part of $\phi_5$):
\begin{dmath*}
Y_{2}\left( \begin{array}{l}- {q}^{+2/3}_{b} {q}^{+2/3}_{g} {\textcolor{blue}{\phi_{5}}}^{-4/3}_{r} \epsilon^{rbg} + {q}^{+2/3}_{g} {q}^{+2/3}_{b} {\textcolor{blue}{\phi_{5}}}^{-4/3}_{r} \epsilon^{rgb} 
\\\\
 + {q}^{-1/3}_{g} {q}^{-1/3}_{b} {\textcolor{blue}{\phi_{5}}}^{+2/3}_{r} \epsilon^{rgb}
- {q}^{-1/3}_{b} {q}^{-1/3}_{g} {\textcolor{blue}{\phi_{5}}}^{+2/3}_{r} \epsilon^{rbg}
\\\\ 
+ \frac{\sqrt{2}}{2}\, {q}^{+2/3}_{b} {q}^{-1/3}_{g} {\textcolor{blue}{\phi_{5}}}^{-1/3}_{r} \epsilon^{rbg}  - \frac{\sqrt{2}}{2}\, {q}^{+2/3}_{g} {q}^{-1/3}_{b} {\textcolor{blue}{\phi_{5}}}^{-1/3}_{r}  \epsilon^{rgb} \\\\  + \frac{\sqrt{2}}{2}\, {q}^{-1/3}_{b} {q}^{+2/3}_{g} {\textcolor{blue}{\phi_{5}}}^{-1/3}_{r} \epsilon^{rbg}     - \frac{\sqrt{2}}{2}\, {q}^{-1/3}_{g} {q}^{+2/3}_{b} {\textcolor{blue}{\phi_{5}}}^{-1/3}_{r}  \epsilon^{rgb}
\end{array} \right) + \mathrm{h.c.} \label{eq:LQ:Y2}
\end{dmath*}
For clarity, we have explicitly included the $SU(3)$ Levi-Civita tensor factors in the above equation. Using the properties of the relevant tensors together with the fermion interchange rules, it follows that the $Y_2$ coupling matrix is antisymmetric in flavor space. Although this result is well known, we provide a brief derivation below for completeness. In this derivation,  ${\textcolor{blue}{\phi_{5}}}$ is written in the adjoint representation. The Yukawa interaction can be explicitly written as:
\begin{align*}
O^{ij}= Y^{ij}_2 Q^{a\alpha r}_i \epsilon_{ad} \epsilon_{\alpha\beta}  Q^{c\beta g}_j  {\textcolor{blue}{\phi^b_{5}}}^d_c   \epsilon_{rgb}
\end{align*}
where $i,j$ denote generation indices, $a,c,d$ are $SU(2)$ indices, $\alpha$ and $\beta$ are Lorentz indices, and $r$, $g$, and $b$ are color indices. Now, we interchange the flavor indices,
\begin{align*}
i\leftrightarrow j:\quad  O^{ji}= Y^{ji}_2 Q^{a\alpha r}_j \epsilon_{ad} \epsilon_{\alpha\beta}  Q^{c\beta g}_i  {\textcolor{blue}{\phi^b_{5}}}^d_c  \epsilon_{rgb} 
\end{align*}
Then we swap the fermions as
\begin{align*}
\psi\leftrightarrow \chi:\quad  O^{ji}= (-1)^1 Y^{ji}_2  Q^{c\beta g}_i  \epsilon_{ad} \epsilon_{\alpha\beta}  Q^{a\alpha r}_j  {\textcolor{blue}{\phi^b_{5}}}^d_c  \epsilon_{rgb}
\end{align*}
moreover, 
\begin{align*}
a\leftrightarrow c, \alpha\leftrightarrow \beta: \quad  O^{ji}&= (-1)^2 Y^{ji}_2  Q^{a\alpha g}_i  \epsilon_{cd} \epsilon_{\alpha\beta}  Q^{c\alpha r}_j  {\textcolor{blue}{\phi^b_{5}}}^d_a  \epsilon_{rgb}
\end{align*}
where the two minus signs arise from
\[
(-1)^2
=
(-1)_{\text{fermion swap}}
(-1)_{\epsilon_{\alpha\beta}}.
\]
Finally,
\begin{align*}
r\leftrightarrow g:\quad  O^{ji}&= (-1)^3  Y^{ji}_2  Q^{a\alpha r}_i  \epsilon_{ad} \epsilon_{\alpha\beta}  Q^{c\alpha g}_j  {\textcolor{blue}{\phi^b_{5}}}^d_c  \epsilon_{rgb}
\end{align*}
shows the origin of the remaining minus sign,
\[
(-1)^3
=
(-1)_{\text{fermion swap}}
(-1)_{\epsilon_{\alpha\beta}} (-1)_{\text{color swap}}.
\]

Therefore, the corresponding Yukawa coupling takes the familiar form
\begin{align*}
Y^{ij}_2 \left(   {\textcolor{blue}{\phi_{5}}}^{-4/3} u_i u_j + {\textcolor{blue}{\phi_{5}}}^{+2/3} d_i d_j - \frac{1}{\sqrt{2}} {\textcolor{blue}{\phi_{5}}}^{-1/3} u_i d_j - \frac{1}{\sqrt{2}} {\textcolor{blue}{\phi_{5}}}^{-1/3} d_i u_j 
\right).
\end{align*}
While this reproduces the known form given in Eq.~(2) of Ref.~\cite{Dorsner:2016wpm}, a careful reader may notice that an additional unphysical phase redefinition, ${\textcolor{blue}{\phi_{5}}}^{\pm 2/3}\to - {\textcolor{blue}{\phi_{5}}}^{\pm 2/3}$ can be performed to reproduce all overall signs appearing in Eq.~(2) of Ref.~\cite{Dorsner:2016wpm}. Since this transformation is merely a field rephasing, it is unphysical and does not affect the canonical normalization of the field $\textcolor{blue}{\phi_{5}}$:
\begin{dmath*}
m^{2}_{6}\left( \begin{array}{l}{\textcolor{blue}{\phi_{5}}}^{-4/3} {{\textcolor{blue}{\phi_{5}}}^{-4/3}}^\dagger + {\textcolor{blue}{\phi_{5}}}^{-1/3} {{\textcolor{blue}{\phi_{5}}}^{-1/3}}^\dagger  + {\textcolor{blue}{\phi_{5}}}^{+2/3} {{\textcolor{blue}{\phi_{5}}}^{+2/3}}^\dagger\end{array} \right)
\end{dmath*}

Let us move to the other Yukawa interaction of $\textcolor{blue}{\phi_{5}}$ LQ:
\begin{dmath*}
Y_{4}\left( \begin{array}{l} {{\textcolor{blue}{\phi_{5}}}^{+2/3,\bar{r}}}^\dagger {l}^{0} {q}^{+2/3}_{r}   + \frac{\sqrt{2}}{2}\, {{\textcolor{blue}{\phi_{5}}}^{-1/3,\bar{r}}}^\dagger {l}^{-} {q}^{+2/3}_{r} \\\\ + \frac{\sqrt{2}}{2}\, {{\textcolor{blue}{\phi_{5}}}^{-1/3,\bar{r}}}^\dagger {l}^{0} {q}^{-1/3}_{r}   + {{\textcolor{blue}{\phi_{5}}}^{-4/3,\bar{r}}}^\dagger {l}^{-} {q}^{-1/3}_{r}\end{array} \right) + \mathrm{h.c.}
\end{dmath*}
Here, $Y_4$ is an arbitrary $3\times 3$ matrix in the flavor space and the above interactions again reproduces Eq.(2) of Ref.~\cite{Dorsner:2016wpm} (as mentioned above, with the unphysical phase rotation ${\textcolor{blue}{\phi_{5}}}^{\pm 2/3}\to - {\textcolor{blue}{\phi_{5}}}^{\pm 2/3}$) 
\begin{align*}
Y^{ij}_4 \left(   {\textcolor{blue}{\phi_{5}}}^{+4/3} e^-_i d_j + {\textcolor{blue}{\phi_{5}}}^{-2/3} \nu_i u_j + \frac{1}{\sqrt{2}} {\textcolor{blue}{\phi_{5}}}^{+1/3} e^-_i u_j + \frac{1}{\sqrt{2}} {\textcolor{blue}{\phi_{5}}}^{+1/3} \nu_i d_j 
\right).
\end{align*}

\subsubsection{Expanded Scalar Sector}
Furthermore, let us analyze a few selected terms from the scalar potential, in particular, terms with coefficients $\mu_8, \lambda_{13, 17, 19, 21}$.  

\begin{pdf}
\begingroup
\allowdisplaybreaks 
\begin{dmath*}
\mu_{8} \, H \textcolor{blue}{\phi_{4}}^\dagger \textcolor{blue}{\phi_{5}}=
\mu_{8}\left( \begin{array}{l}- \sqrt{2}\, {H}^{0} {{\textcolor{blue}{\phi_{4}}}^{+2/3,\bar{r}}}^\dagger {\textcolor{blue}{\phi_{5}}}^{+2/3}_{r} - \sqrt{2}\, {H}^{0} {{\textcolor{blue}{\phi_{4}}}^{+2/3,\bar{g}}}^\dagger {\textcolor{blue}{\phi_{5}}}^{+2/3}_{g} \\\\ - \sqrt{2}\, {H}^{0} {{\textcolor{blue}{\phi_{4}}}^{+2/3,\bar{b}}}^\dagger {\textcolor{blue}{\phi_{5}}}^{+2/3}_{b} - {H}^{0} {{\textcolor{blue}{\phi_{4}}}^{-1/3,\bar{r}}}^\dagger {\textcolor{blue}{\phi_{5}}}^{-1/3}_{r} \\\\ - {H}^{0} {{\textcolor{blue}{\phi_{4}}}^{-1/3,\bar{g}}}^\dagger {\textcolor{blue}{\phi_{5}}}^{-1/3}_{g} - {H}^{0} {{\textcolor{blue}{\phi_{4}}}^{-1/3,\bar{b}}}^\dagger {\textcolor{blue}{\phi_{5}}}^{-1/3}_{b} \\\\ + {H}^{+} {{\textcolor{blue}{\phi_{4}}}^{+2/3,\bar{r}}}^\dagger {\textcolor{blue}{\phi_{5}}}^{-1/3}_{r} + {H}^{+} {{\textcolor{blue}{\phi_{4}}}^{+2/3,\bar{g}}}^\dagger {\textcolor{blue}{\phi_{5}}}^{-1/3}_{g} \\\\ + {H}^{+} {{\textcolor{blue}{\phi_{4}}}^{+2/3,\bar{b}}}^\dagger {\textcolor{blue}{\phi_{5}}}^{-1/3}_{b} + \sqrt{2}\, {H}^{+} {{\textcolor{blue}{\phi_{4}}}^{-1/3,\bar{r}}}^\dagger {\textcolor{blue}{\phi_{5}}}^{-4/3}_{r} \\\\ + \sqrt{2}\, {H}^{+} {{\textcolor{blue}{\phi_{4}}}^{-1/3,\bar{g}}}^\dagger {\textcolor{blue}{\phi_{5}}}^{-4/3}_{g} + \sqrt{2}\, {H}^{+} {{\textcolor{blue}{\phi_{4}}}^{-1/3,\bar{b}}}^\dagger {\textcolor{blue}{\phi_{5}}}^{-4/3}_{b}\end{array} \right) 
\end{dmath*}

\begin{dmath*}
\lambda_{13} \, H H \textcolor{blue}{\phi_{3}}^\dagger \textcolor{blue}{\phi_{4}} =
\lambda_{13}\left( \begin{array}{l}- {H}^{0} {H}^{0} {{\textcolor{blue}{\phi_{3}}}^{+2/3,\bar{r}}}^\dagger {\textcolor{blue}{\phi_{4}}}^{+2/3}_{r} - {H}^{0} {H}^{0} {{\textcolor{blue}{\phi_{3}}}^{+2/3,\bar{g}}}^\dagger {\textcolor{blue}{\phi_{4}}}^{+2/3}_{g} \\\\ - {H}^{0} {H}^{0} {{\textcolor{blue}{\phi_{3}}}^{+2/3,\bar{b}}}^\dagger {\textcolor{blue}{\phi_{4}}}^{+2/3}_{b} - {H}^{0} {H}^{+} {{\textcolor{blue}{\phi_{3}}}^{+5/3,\bar{r}}}^\dagger {\textcolor{blue}{\phi_{4}}}^{+2/3}_{r} \\\\ - {H}^{0} {H}^{+} {{\textcolor{blue}{\phi_{3}}}^{+5/3,\bar{g}}}^\dagger {\textcolor{blue}{\phi_{4}}}^{+2/3}_{g} - {H}^{0} {H}^{+} {{\textcolor{blue}{\phi_{3}}}^{+5/3,\bar{b}}}^\dagger {\textcolor{blue}{\phi_{4}}}^{+2/3}_{b} \\\\ + {H}^{0} {H}^{+} {{\textcolor{blue}{\phi_{3}}}^{+2/3,\bar{r}}}^\dagger {\textcolor{blue}{\phi_{4}}}^{-1/3}_{r} + {H}^{0} {H}^{+} {{\textcolor{blue}{\phi_{3}}}^{+2/3,\bar{g}}}^\dagger {\textcolor{blue}{\phi_{4}}}^{-1/3}_{g} \\\\ + {H}^{0} {H}^{+} {{\textcolor{blue}{\phi_{3}}}^{+2/3,\bar{b}}}^\dagger {\textcolor{blue}{\phi_{4}}}^{-1/3}_{b} + {H}^{+} {H}^{+} {{\textcolor{blue}{\phi_{3}}}^{+5/3,\bar{r}}}^\dagger {\textcolor{blue}{\phi_{4}}}^{-1/3}_{r} \\\\ + {H}^{+} {H}^{+} {{\textcolor{blue}{\phi_{3}}}^{+5/3,\bar{g}}}^\dagger {\textcolor{blue}{\phi_{4}}}^{-1/3}_{g} + {H}^{+} {H}^{+} {{\textcolor{blue}{\phi_{3}}}^{+5/3,\bar{b}}}^\dagger {\textcolor{blue}{\phi_{4}}}^{-1/3}_{b}\end{array} \right) 
\end{dmath*}

\begin{dmath*}
\lambda_{17,c1} \, H H^\dagger \textcolor{blue}{\phi_{3}} \textcolor{blue}{\phi_{3}}^\dagger = 
\lambda_{17,c1}^{(1)}\left( \begin{array}{l}{H}^{0} {{H}^{+}}^\dagger {\textcolor{blue}{\phi_{3}}}^{+5/3}_{r} {{\textcolor{blue}{\phi_{3}}}^{+2/3,\bar{r}}}^\dagger + {H}^{0} {{H}^{+}}^\dagger {\textcolor{blue}{\phi_{3}}}^{+5/3}_{g} {{\textcolor{blue}{\phi_{3}}}^{+2/3,\bar{g}}}^\dagger \\\\ + {H}^{0} {{H}^{+}}^\dagger {\textcolor{blue}{\phi_{3}}}^{+5/3}_{b} {{\textcolor{blue}{\phi_{3}}}^{+2/3,\bar{b}}}^\dagger + {H}^{0} {{H}^{0}}^\dagger {\textcolor{blue}{\phi_{3}}}^{+2/3}_{r} {{\textcolor{blue}{\phi_{3}}}^{+2/3,\bar{r}}}^\dagger \\\\ + {H}^{0} {{H}^{0}}^\dagger {\textcolor{blue}{\phi_{3}}}^{+2/3}_{g} {{\textcolor{blue}{\phi_{3}}}^{+2/3,\bar{g}}}^\dagger + {H}^{0} {{H}^{0}}^\dagger {\textcolor{blue}{\phi_{3}}}^{+2/3}_{b} {{\textcolor{blue}{\phi_{3}}}^{+2/3,\bar{b}}}^\dagger \\\\ + {H}^{+} {{H}^{+}}^\dagger {\textcolor{blue}{\phi_{3}}}^{+5/3}_{r} {{\textcolor{blue}{\phi_{3}}}^{+5/3,\bar{r}}}^\dagger + {H}^{+} {{H}^{+}}^\dagger {\textcolor{blue}{\phi_{3}}}^{+5/3}_{g} {{\textcolor{blue}{\phi_{3}}}^{+5/3,\bar{g}}}^\dagger \\\\ + {H}^{+} {{H}^{+}}^\dagger {\textcolor{blue}{\phi_{3}}}^{+5/3}_{b} {{\textcolor{blue}{\phi_{3}}}^{+5/3,\bar{b}}}^\dagger + {H}^{+} {{H}^{0}}^\dagger {\textcolor{blue}{\phi_{3}}}^{+2/3}_{r} {{\textcolor{blue}{\phi_{3}}}^{+5/3,\bar{r}}}^\dagger \\\\ + {H}^{+} {{H}^{0}}^\dagger {\textcolor{blue}{\phi_{3}}}^{+2/3}_{g} {{\textcolor{blue}{\phi_{3}}}^{+5/3,\bar{g}}}^\dagger + {H}^{+} {{H}^{0}}^\dagger {\textcolor{blue}{\phi_{3}}}^{+2/3}_{b} {{\textcolor{blue}{\phi_{3}}}^{+5/3,\bar{b}}}^\dagger\end{array} \right)
\end{dmath*}

\begin{dmath*}
\lambda_{17,c2} \, H H^\dagger \textcolor{blue}{\phi_{3}} \textcolor{blue}{\phi_{3}}^\dagger = \lambda_{17,c2}^{(2)}\left( \begin{array}{l}- {H}^{0} {{H}^{+}}^\dagger {\textcolor{blue}{\phi_{3}}}^{+5/3}_{r} {{\textcolor{blue}{\phi_{3}}}^{+2/3,\bar{r}}}^\dagger - {H}^{0} {{H}^{+}}^\dagger {\textcolor{blue}{\phi_{3}}}^{+5/3}_{g} {{\textcolor{blue}{\phi_{3}}}^{+2/3,\bar{g}}}^\dagger \\\\ - {H}^{0} {{H}^{+}}^\dagger {\textcolor{blue}{\phi_{3}}}^{+5/3}_{b} {{\textcolor{blue}{\phi_{3}}}^{+2/3,\bar{b}}}^\dagger + {H}^{0} {{H}^{0}}^\dagger {\textcolor{blue}{\phi_{3}}}^{+5/3}_{r} {{\textcolor{blue}{\phi_{3}}}^{+5/3,\bar{r}}}^\dagger \\\\ + {H}^{0} {{H}^{0}}^\dagger {\textcolor{blue}{\phi_{3}}}^{+5/3}_{g} {{\textcolor{blue}{\phi_{3}}}^{+5/3,\bar{g}}}^\dagger + {H}^{0} {{H}^{0}}^\dagger {\textcolor{blue}{\phi_{3}}}^{+5/3}_{b} {{\textcolor{blue}{\phi_{3}}}^{+5/3,\bar{b}}}^\dagger \\\\ + {H}^{+} {{H}^{+}}^\dagger {\textcolor{blue}{\phi_{3}}}^{+2/3}_{r} {{\textcolor{blue}{\phi_{3}}}^{+2/3,\bar{r}}}^\dagger + {H}^{+} {{H}^{+}}^\dagger {\textcolor{blue}{\phi_{3}}}^{+2/3}_{g} {{\textcolor{blue}{\phi_{3}}}^{+2/3,\bar{g}}}^\dagger \\\\ + {H}^{+} {{H}^{+}}^\dagger {\textcolor{blue}{\phi_{3}}}^{+2/3}_{b} {{\textcolor{blue}{\phi_{3}}}^{+2/3,\bar{b}}}^\dagger - {H}^{+} {{H}^{0}}^\dagger {\textcolor{blue}{\phi_{3}}}^{+2/3}_{r} {{\textcolor{blue}{\phi_{3}}}^{+5/3,\bar{r}}}^\dagger \\\\ - {H}^{+} {{H}^{0}}^\dagger {\textcolor{blue}{\phi_{3}}}^{+2/3}_{g} {{\textcolor{blue}{\phi_{3}}}^{+5/3,\bar{g}}}^\dagger - {H}^{+} {{H}^{0}}^\dagger {\textcolor{blue}{\phi_{3}}}^{+2/3}_{b} {{\textcolor{blue}{\phi_{3}}}^{+5/3,\bar{b}}}^\dagger\end{array} \right)
\end{dmath*}

\begin{dmath*}
\lambda_{19,c1} \, H H^\dagger \textcolor{blue}{\phi_{4}} \textcolor{blue}{\phi_{4}}^\dagger = \lambda_{19,c1}^{(1)}\left( \begin{array}{l}{H}^{0} {{H}^{+}}^\dagger {\textcolor{blue}{\phi_{4}}}^{+2/3}_{r} {{\textcolor{blue}{\phi_{4}}}^{-1/3,\bar{r}}}^\dagger + {H}^{0} {{H}^{+}}^\dagger {\textcolor{blue}{\phi_{4}}}^{+2/3}_{g} {{\textcolor{blue}{\phi_{4}}}^{-1/3,\bar{g}}}^\dagger \\\\ + {H}^{0} {{H}^{+}}^\dagger {\textcolor{blue}{\phi_{4}}}^{+2/3}_{b} {{\textcolor{blue}{\phi_{4}}}^{-1/3,\bar{b}}}^\dagger + {H}^{0} {{H}^{0}}^\dagger {\textcolor{blue}{\phi_{4}}}^{-1/3}_{r} {{\textcolor{blue}{\phi_{4}}}^{-1/3,\bar{r}}}^\dagger \\\\ + {H}^{0} {{H}^{0}}^\dagger {\textcolor{blue}{\phi_{4}}}^{-1/3}_{g} {{\textcolor{blue}{\phi_{4}}}^{-1/3,\bar{g}}}^\dagger + {H}^{0} {{H}^{0}}^\dagger {\textcolor{blue}{\phi_{4}}}^{-1/3}_{b} {{\textcolor{blue}{\phi_{4}}}^{-1/3,\bar{b}}}^\dagger \\\\ + {H}^{+} {{H}^{+}}^\dagger {\textcolor{blue}{\phi_{4}}}^{+2/3}_{r} {{\textcolor{blue}{\phi_{4}}}^{+2/3,\bar{r}}}^\dagger + {H}^{+} {{H}^{+}}^\dagger {\textcolor{blue}{\phi_{4}}}^{+2/3}_{g} {{\textcolor{blue}{\phi_{4}}}^{+2/3,\bar{g}}}^\dagger \\\\ + {H}^{+} {{H}^{+}}^\dagger {\textcolor{blue}{\phi_{4}}}^{+2/3}_{b} {{\textcolor{blue}{\phi_{4}}}^{+2/3,\bar{b}}}^\dagger + {H}^{+} {{H}^{0}}^\dagger {\textcolor{blue}{\phi_{4}}}^{-1/3}_{r} {{\textcolor{blue}{\phi_{4}}}^{+2/3,\bar{r}}}^\dagger \\\\ + {H}^{+} {{H}^{0}}^\dagger {\textcolor{blue}{\phi_{4}}}^{-1/3}_{g} {{\textcolor{blue}{\phi_{4}}}^{+2/3,\bar{g}}}^\dagger + {H}^{+} {{H}^{0}}^\dagger {\textcolor{blue}{\phi_{4}}}^{-1/3}_{b} {{\textcolor{blue}{\phi_{4}}}^{+2/3,\bar{b}}}^\dagger\end{array} \right)
\end{dmath*}

\begin{dmath*}
\lambda_{19,c2} \, H H^\dagger \textcolor{blue}{\phi_{4}} \textcolor{blue}{\phi_{4}}^\dagger= \lambda_{19,c2}^{(2)}\left( \begin{array}{l}- {H}^{0} {{H}^{+}}^\dagger {\textcolor{blue}{\phi_{4}}}^{+2/3}_{r} {{\textcolor{blue}{\phi_{4}}}^{-1/3,\bar{r}}}^\dagger - {H}^{0} {{H}^{+}}^\dagger {\textcolor{blue}{\phi_{4}}}^{+2/3}_{g} {{\textcolor{blue}{\phi_{4}}}^{-1/3,\bar{g}}}^\dagger \\\\ - {H}^{0} {{H}^{+}}^\dagger {\textcolor{blue}{\phi_{4}}}^{+2/3}_{b} {{\textcolor{blue}{\phi_{4}}}^{-1/3,\bar{b}}}^\dagger + {H}^{0} {{H}^{0}}^\dagger {\textcolor{blue}{\phi_{4}}}^{+2/3}_{r} {{\textcolor{blue}{\phi_{4}}}^{+2/3,\bar{r}}}^\dagger \\\\ + {H}^{0} {{H}^{0}}^\dagger {\textcolor{blue}{\phi_{4}}}^{+2/3}_{g} {{\textcolor{blue}{\phi_{4}}}^{+2/3,\bar{g}}}^\dagger + {H}^{0} {{H}^{0}}^\dagger {\textcolor{blue}{\phi_{4}}}^{+2/3}_{b} {{\textcolor{blue}{\phi_{4}}}^{+2/3,\bar{b}}}^\dagger \\\\ + {H}^{+} {{H}^{+}}^\dagger {\textcolor{blue}{\phi_{4}}}^{-1/3}_{r} {{\textcolor{blue}{\phi_{4}}}^{-1/3,\bar{r}}}^\dagger + {H}^{+} {{H}^{+}}^\dagger {\textcolor{blue}{\phi_{4}}}^{-1/3}_{g} {{\textcolor{blue}{\phi_{4}}}^{-1/3,\bar{g}}}^\dagger \\\\ + {H}^{+} {{H}^{+}}^\dagger {\textcolor{blue}{\phi_{4}}}^{-1/3}_{b} {{\textcolor{blue}{\phi_{4}}}^{-1/3,\bar{b}}}^\dagger - {H}^{+} {{H}^{0}}^\dagger {\textcolor{blue}{\phi_{4}}}^{-1/3}_{r} {{\textcolor{blue}{\phi_{4}}}^{+2/3,\bar{r}}}^\dagger \\\\ - {H}^{+} {{H}^{0}}^\dagger {\textcolor{blue}{\phi_{4}}}^{-1/3}_{g} {{\textcolor{blue}{\phi_{4}}}^{+2/3,\bar{g}}}^\dagger - {H}^{+} {{H}^{0}}^\dagger {\textcolor{blue}{\phi_{4}}}^{-1/3}_{b} {{\textcolor{blue}{\phi_{4}}}^{+2/3,\bar{b}}}^\dagger\end{array} \right)
\end{dmath*}

\begin{dmath*}
\lambda_{21,c1} \, H H^\dagger \textcolor{blue}{\phi_{5}} \textcolor{blue}{\phi_{5}}^\dagger = \lambda_{21,c1}^{(1)}\left( \begin{array}{l}\sqrt{2}\, {H}^{0} {{H}^{+}}^\dagger {\textcolor{blue}{\phi_{5}}}^{-1/3}_{r} {{\textcolor{blue}{\phi_{5}}}^{-4/3,\bar{r}}}^\dagger + \sqrt{2}\, {H}^{0} {{H}^{+}}^\dagger {\textcolor{blue}{\phi_{5}}}^{-1/3}_{g} {{\textcolor{blue}{\phi_{5}}}^{-4/3,\bar{g}}}^\dagger \\\\ + \sqrt{2}\, {H}^{0} {{H}^{+}}^\dagger {\textcolor{blue}{\phi_{5}}}^{-1/3}_{b} {{\textcolor{blue}{\phi_{5}}}^{-4/3,\bar{b}}}^\dagger + \sqrt{2}\, {H}^{0} {{H}^{+}}^\dagger {\textcolor{blue}{\phi_{5}}}^{+2/3}_{r} {{\textcolor{blue}{\phi_{5}}}^{-1/3,\bar{r}}}^\dagger \\\\ + \sqrt{2}\, {H}^{0} {{H}^{+}}^\dagger {\textcolor{blue}{\phi_{5}}}^{+2/3}_{g} {{\textcolor{blue}{\phi_{5}}}^{-1/3,\bar{g}}}^\dagger + \sqrt{2}\, {H}^{0} {{H}^{+}}^\dagger {\textcolor{blue}{\phi_{5}}}^{+2/3}_{b} {{\textcolor{blue}{\phi_{5}}}^{-1/3,\bar{b}}}^\dagger \\\\ + {H}^{0} {{H}^{0}}^\dagger {\textcolor{blue}{\phi_{5}}}^{-4/3}_{r} {{\textcolor{blue}{\phi_{5}}}^{-4/3,\bar{r}}}^\dagger + {H}^{0} {{H}^{0}}^\dagger {\textcolor{blue}{\phi_{5}}}^{-4/3}_{g} {{\textcolor{blue}{\phi_{5}}}^{-4/3,\bar{g}}}^\dagger \\\\ + {H}^{0} {{H}^{0}}^\dagger {\textcolor{blue}{\phi_{5}}}^{-4/3}_{b} {{\textcolor{blue}{\phi_{5}}}^{-4/3,\bar{b}}}^\dagger - {H}^{0} {{H}^{0}}^\dagger {\textcolor{blue}{\phi_{5}}}^{+2/3}_{r} {{\textcolor{blue}{\phi_{5}}}^{+2/3,\bar{r}}}^\dagger \\\\ - {H}^{0} {{H}^{0}}^\dagger {\textcolor{blue}{\phi_{5}}}^{+2/3}_{g} {{\textcolor{blue}{\phi_{5}}}^{+2/3,\bar{g}}}^\dagger - {H}^{0} {{H}^{0}}^\dagger {\textcolor{blue}{\phi_{5}}}^{+2/3}_{b} {{\textcolor{blue}{\phi_{5}}}^{+2/3,\bar{b}}}^\dagger \\\\ - {H}^{+} {{H}^{+}}^\dagger {\textcolor{blue}{\phi_{5}}}^{-4/3}_{r} {{\textcolor{blue}{\phi_{5}}}^{-4/3,\bar{r}}}^\dagger - {H}^{+} {{H}^{+}}^\dagger {\textcolor{blue}{\phi_{5}}}^{-4/3}_{g} {{\textcolor{blue}{\phi_{5}}}^{-4/3,\bar{g}}}^\dagger \\\\ - {H}^{+} {{H}^{+}}^\dagger {\textcolor{blue}{\phi_{5}}}^{-4/3}_{b} {{\textcolor{blue}{\phi_{5}}}^{-4/3,\bar{b}}}^\dagger + {H}^{+} {{H}^{+}}^\dagger {\textcolor{blue}{\phi_{5}}}^{+2/3}_{r} {{\textcolor{blue}{\phi_{5}}}^{+2/3,\bar{r}}}^\dagger \\\\ + {H}^{+} {{H}^{+}}^\dagger {\textcolor{blue}{\phi_{5}}}^{+2/3}_{g} {{\textcolor{blue}{\phi_{5}}}^{+2/3,\bar{g}}}^\dagger + {H}^{+} {{H}^{+}}^\dagger {\textcolor{blue}{\phi_{5}}}^{+2/3}_{b} {{\textcolor{blue}{\phi_{5}}}^{+2/3,\bar{b}}}^\dagger \\\\ + \sqrt{2}\, {H}^{+} {{H}^{0}}^\dagger {\textcolor{blue}{\phi_{5}}}^{-4/3}_{r} {{\textcolor{blue}{\phi_{5}}}^{-1/3,\bar{r}}}^\dagger + \sqrt{2}\, {H}^{+} {{H}^{0}}^\dagger {\textcolor{blue}{\phi_{5}}}^{-4/3}_{g} {{\textcolor{blue}{\phi_{5}}}^{-1/3,\bar{g}}}^\dagger \\\\ + \sqrt{2}\, {H}^{+} {{H}^{0}}^\dagger {\textcolor{blue}{\phi_{5}}}^{-4/3}_{b} {{\textcolor{blue}{\phi_{5}}}^{-1/3,\bar{b}}}^\dagger + \sqrt{2}\, {H}^{+} {{H}^{0}}^\dagger {\textcolor{blue}{\phi_{5}}}^{-1/3}_{r} {{\textcolor{blue}{\phi_{5}}}^{+2/3,\bar{r}}}^\dagger \\\\ + \sqrt{2}\, {H}^{+} {{H}^{0}}^\dagger {\textcolor{blue}{\phi_{5}}}^{-1/3}_{g} {{\textcolor{blue}{\phi_{5}}}^{+2/3,\bar{g}}}^\dagger + \sqrt{2}\, {H}^{+} {{H}^{0}}^\dagger {\textcolor{blue}{\phi_{5}}}^{-1/3}_{b} {{\textcolor{blue}{\phi_{5}}}^{+2/3,\bar{b}}}^\dagger\end{array} \right)
\end{dmath*}

\begin{dmath*}
\lambda_{21,c2} \, H H^\dagger \textcolor{blue}{\phi_{5}} \textcolor{blue}{\phi_{5}}^\dagger= \lambda_{21,c2}^{(2)}\left( \begin{array}{l}- \sqrt{2}\, {H}^{0} {{H}^{+}}^\dagger {\textcolor{blue}{\phi_{5}}}^{-1/3}_{r} {{\textcolor{blue}{\phi_{5}}}^{-4/3,\bar{r}}}^\dagger - \sqrt{2}\, {H}^{0} {{H}^{+}}^\dagger {\textcolor{blue}{\phi_{5}}}^{-1/3}_{g} {{\textcolor{blue}{\phi_{5}}}^{-4/3,\bar{g}}}^\dagger \\\\ - \sqrt{2}\, {H}^{0} {{H}^{+}}^\dagger {\textcolor{blue}{\phi_{5}}}^{-1/3}_{b} {{\textcolor{blue}{\phi_{5}}}^{-4/3,\bar{b}}}^\dagger - \sqrt{2}\, {H}^{0} {{H}^{+}}^\dagger {\textcolor{blue}{\phi_{5}}}^{+2/3}_{r} {{\textcolor{blue}{\phi_{5}}}^{-1/3,\bar{r}}}^\dagger \\\\ - \sqrt{2}\, {H}^{0} {{H}^{+}}^\dagger {\textcolor{blue}{\phi_{5}}}^{+2/3}_{g} {{\textcolor{blue}{\phi_{5}}}^{-1/3,\bar{g}}}^\dagger - \sqrt{2}\, {H}^{0} {{H}^{+}}^\dagger {\textcolor{blue}{\phi_{5}}}^{+2/3}_{b} {{\textcolor{blue}{\phi_{5}}}^{-1/3,\bar{b}}}^\dagger \\\\ + {H}^{0} {{H}^{0}}^\dagger {\textcolor{blue}{\phi_{5}}}^{-1/3}_{r} {{\textcolor{blue}{\phi_{5}}}^{-1/3,\bar{r}}}^\dagger + {H}^{0} {{H}^{0}}^\dagger {\textcolor{blue}{\phi_{5}}}^{-1/3}_{g} {{\textcolor{blue}{\phi_{5}}}^{-1/3,\bar{g}}}^\dagger \\\\ + {H}^{0} {{H}^{0}}^\dagger {\textcolor{blue}{\phi_{5}}}^{-1/3}_{b} {{\textcolor{blue}{\phi_{5}}}^{-1/3,\bar{b}}}^\dagger + 2\, {H}^{0} {{H}^{0}}^\dagger {\textcolor{blue}{\phi_{5}}}^{+2/3}_{r} {{\textcolor{blue}{\phi_{5}}}^{+2/3,\bar{r}}}^\dagger \\\\ + 2\, {H}^{0} {{H}^{0}}^\dagger {\textcolor{blue}{\phi_{5}}}^{+2/3}_{g} {{\textcolor{blue}{\phi_{5}}}^{+2/3,\bar{g}}}^\dagger + 2\, {H}^{0} {{H}^{0}}^\dagger {\textcolor{blue}{\phi_{5}}}^{+2/3}_{b} {{\textcolor{blue}{\phi_{5}}}^{+2/3,\bar{b}}}^\dagger \\\\ + 2\, {H}^{+} {{H}^{+}}^\dagger {\textcolor{blue}{\phi_{5}}}^{-4/3}_{r} {{\textcolor{blue}{\phi_{5}}}^{-4/3,\bar{r}}}^\dagger + 2\, {H}^{+} {{H}^{+}}^\dagger {\textcolor{blue}{\phi_{5}}}^{-4/3}_{g} {{\textcolor{blue}{\phi_{5}}}^{-4/3,\bar{g}}}^\dagger \\\\ + 2\, {H}^{+} {{H}^{+}}^\dagger {\textcolor{blue}{\phi_{5}}}^{-4/3}_{b} {{\textcolor{blue}{\phi_{5}}}^{-4/3,\bar{b}}}^\dagger + {H}^{+} {{H}^{+}}^\dagger {\textcolor{blue}{\phi_{5}}}^{-1/3}_{r} {{\textcolor{blue}{\phi_{5}}}^{-1/3,\bar{r}}}^\dagger \\\\ + {H}^{+} {{H}^{+}}^\dagger {\textcolor{blue}{\phi_{5}}}^{-1/3}_{g} {{\textcolor{blue}{\phi_{5}}}^{-1/3,\bar{g}}}^\dagger + {H}^{+} {{H}^{+}}^\dagger {\textcolor{blue}{\phi_{5}}}^{-1/3}_{b} {{\textcolor{blue}{\phi_{5}}}^{-1/3,\bar{b}}}^\dagger \\\\ - \sqrt{2}\, {H}^{+} {{H}^{0}}^\dagger {\textcolor{blue}{\phi_{5}}}^{-4/3}_{r} {{\textcolor{blue}{\phi_{5}}}^{-1/3,\bar{r}}}^\dagger - \sqrt{2}\, {H}^{+} {{H}^{0}}^\dagger {\textcolor{blue}{\phi_{5}}}^{-4/3}_{g} {{\textcolor{blue}{\phi_{5}}}^{-1/3,\bar{g}}}^\dagger \\\\ - \sqrt{2}\, {H}^{+} {{H}^{0}}^\dagger {\textcolor{blue}{\phi_{5}}}^{-4/3}_{b} {{\textcolor{blue}{\phi_{5}}}^{-1/3,\bar{b}}}^\dagger - \sqrt{2}\, {H}^{+} {{H}^{0}}^\dagger {\textcolor{blue}{\phi_{5}}}^{-1/3}_{r} {{\textcolor{blue}{\phi_{5}}}^{+2/3,\bar{r}}}^\dagger \\\\ - \sqrt{2}\, {H}^{+} {{H}^{0}}^\dagger {\textcolor{blue}{\phi_{5}}}^{-1/3}_{g} {{\textcolor{blue}{\phi_{5}}}^{+2/3,\bar{g}}}^\dagger - \sqrt{2}\, {H}^{+} {{H}^{0}}^\dagger {\textcolor{blue}{\phi_{5}}}^{-1/3}_{b} {{\textcolor{blue}{\phi_{5}}}^{+2/3,\bar{b}}}^\dagger\end{array} \right)
\end{dmath*}
\endgroup
\end{pdf}

These selected terms can be directly mapped onto the notation used in Ref.~\cite{Crivellin:2021ejk} via the following identifications:
\begin{align}
A_{\widetilde 2 3} &= \sqrt{2}\mu_8^* \label{eq:01} \\
Y_{2 \widetilde 2} &= \lambda_{13}^* \\
Y_3 &= \lambda_{17,c_1} \\
Y_{22}  &= \lambda_{17,c_2}-\lambda_{17,c_1} \\
Y_4 &= \lambda_{19,c_1} \\
Y_{\widetilde 2\widetilde 2} &= \lambda_{19,c_2} - \lambda_{19,c_1}\\
Y_{33} &=\lambda_{21,c_2} - \lambda_{21,c_1} \\
Y_5 &= \lambda_{21,c_2} \label{eq:02}
\end{align}
It should be emphasized that the couplings appearing on the left-hand side of Eqs.~\eqref{eq:01}--\eqref{eq:02} follow the notation of Ref.~\cite{Crivellin:2021ejk} and should not be confused with the Yukawa coupling matrices introduced elsewhere in this work. Furthermore, when performing this comparison, the field redefinition ${\textcolor{blue}{\phi_{5}}}^{\pm 2/3}\to - {\textcolor{blue}{\phi_{5}}}^{\pm 2/3}$ should be taken into account, as stated above.

\subsection{Getting the Mass Matrices}
The user can then request the package to generate the mass matrices. 

\begin{user}
    generate mass matrices
\end{user}

\begin{agent}
\begin{verbatim}
Generated report for 'SM+phi_1+phi_2+phi_3+phi_4+phi_5' 
with field content (3,1,-1/3), (3,1,-4/3), (3,2,7/6), (3,2,1/6), (3,3,-1/3). 
TeX: /home/user/Downloads/bsm_agent/output/SM_phi_1_phi_2_phi_3_phi_4_phi_5.tex. 
PDF: /home/user/Downloads/bsm_agent/output/SM_phi_1_phi_2_phi_3_phi_4_phi_5.pdf. 
[22.9s]
\end{verbatim}
\end{agent}

\begin{pdf}
\centering
\includegraphics[width=0.55\textwidth]{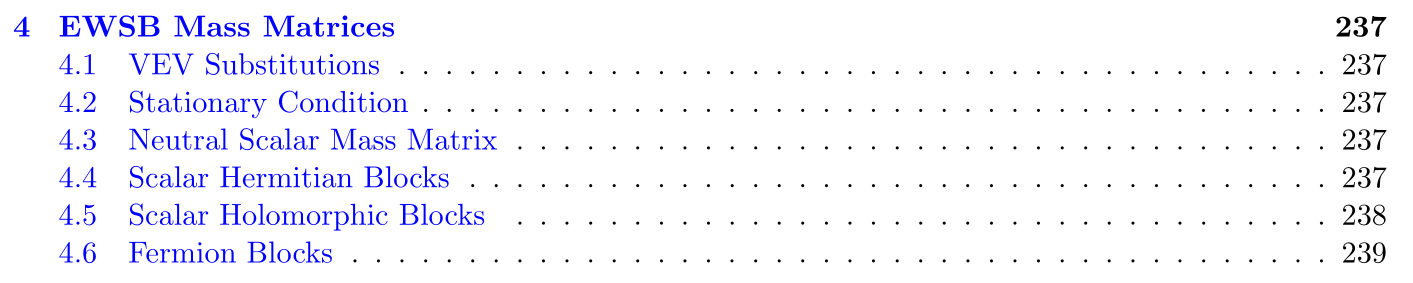}
\end{pdf}

Let us first look at the mass matrix for the fields $\phi^{\pm 2/3}$:
\begin{pdf}
\[
\begin{array}{l}
M^2_{10}[\left({\textcolor{blue}{\phi_{3}}}^{+2/3}_{r}\right)^\dagger, \left({\textcolor{blue}{\phi_{4}}}^{+2/3}_{r}\right)^\dagger, \left({\textcolor{blue}{\phi_{5}}}^{+2/3}_{r}\right)^\dagger;{\textcolor{blue}{\phi_{3}}}^{+2/3}_{r}, {\textcolor{blue}{\phi_{4}}}^{+2/3}_{r}, {\textcolor{blue}{\phi_{5}}}^{+2/3}_{r}] = \\
\left[\begin{matrix}\frac{\lambda_{17,c1} v^{2}}{2} + m^{2}_{4} & - \frac{\lambda_{13} v^{2}}{2} & 0\\- \frac{v^{2} \overline{\lambda_{13}}}{2} & \frac{\lambda_{19,c2} v^{2}}{2} + m^{2}_{5} & - \mu_{8} v\\0 & - v \overline{\mu_{8}} & - \frac{\lambda_{21,c1} v^{2}}{2} + \lambda_{21,c2} v^{2} + m^{2}_{6}\end{matrix}\right]
\end{array}
\]    
\end{pdf}
which, as expected, agrees with the result of Ref.~\cite{Crivellin:2021ejk} once the field redefinition  ${\textcolor{blue}{\phi_{5}}}^{\pm 2/3}\to - {\textcolor{blue}{\phi_{5}}}^{\pm 2/3}$ is taken into account. In these mass matrices, for example, $\overline{\lambda_{13}}$ refers to  $\lambda_{13}^*$, and so on.

Similarly, the other mass matrices, reported by the package are as follows. For the $\phi^{\pm 1/3}$ fields:
\begin{pdf}
\[
\begin{array}{l}
M^2_{4}[\left({\textcolor{blue}{\phi_{1}}}^{-1/3}_{r}\right)^\dagger, \left({\textcolor{blue}{\phi_{4}}}^{-1/3}_{r}\right)^\dagger, \left({\textcolor{blue}{\phi_{5}}}^{-1/3}_{r}\right)^\dagger;{\textcolor{blue}{\phi_{1}}}^{-1/3}_{r}, {\textcolor{blue}{\phi_{4}}}^{-1/3}_{r}, {\textcolor{blue}{\phi_{5}}}^{-1/3}_{r}] = \\
\left[\begin{matrix}\frac{\lambda_{14} v^{2}}{2} + m^{2}_{2} & \frac{\sqrt{2} v \overline{\mu_{7}}}{2} & - \frac{v^{2} \overline{\lambda_{15}}}{2}\\\frac{\sqrt{2} \mu_{7} v}{2} & \frac{\lambda_{19,c1} v^{2}}{2} + m^{2}_{5} & - \frac{\sqrt{2} \mu_{8} v}{2}\\- \frac{\lambda_{15} v^{2}}{2} & - \frac{\sqrt{2} v \overline{\mu_{8}}}{2} & \frac{\lambda_{21,c2} v^{2}}{2} + m^{2}_{6}\end{matrix}\right]
\end{array}
\]    
\end{pdf}
The mass matrix for the $\phi^{\pm 4/3}$ fields, as obtained by the package, is
\begin{pdf}
\[
\begin{array}{l}
M^2_{7}[\left({\textcolor{blue}{\phi_{2}}}^{-4/3}_{r}\right)^\dagger, \left({\textcolor{blue}{\phi_{5}}}^{-4/3}_{r}\right)^\dagger;{\textcolor{blue}{\phi_{2}}}^{-4/3}_{r}, {\textcolor{blue}{\phi_{5}}}^{-4/3}_{r}] = \\
\left[\begin{matrix}\frac{\lambda_{16} v^{2}}{2} + m^{2}_{3} & \frac{v^{2} \overline{\lambda_{12}}}{2}\\\frac{\lambda_{12} v^{2}}{2} & \frac{\lambda_{21,c1} v^{2}}{2} + m^{2}_{6}\end{matrix}\right]
\end{array}
\]    
\end{pdf}
Finally, for the $\phi^{\pm 5/3}$ fields, the package computes the following mass matrix:
\begin{pdf}
\[
\begin{array}{l}
M^2_{13}[\left({\textcolor{blue}{\phi_{3}}}^{+5/3}_{r}\right)^\dagger;{\textcolor{blue}{\phi_{3}}}^{+5/3}_{r}] = \\
\left[\begin{matrix}\frac{\lambda_{17,c2} v^{2}}{2} + m^{2}_{4}\end{matrix}\right]
\end{array}
\]    
\end{pdf}
The results can be directly compared with those of Ref.~\cite{Crivellin:2021ejk} using the additional mapping as given below:
\begin{align}
A_{1\widetilde 2} &= \mu_7^* \label{eq:03} \\
Y_1 &= \lambda_{14} \\
Y_2 &= \lambda_{16} \\
Y_{\widetilde 13} &= -\lambda_{12}^*\\
Y_{13} &=\sqrt{2} \lambda_{15}  \label{eq:04}
\end{align}

\subsection{Model construction at one go}
In the above subsection, the model was constructed by adding the five BSM fields sequentially, one at a time. However, a user may instead prefer to add all BSM fields simultaneously in a single request.
\begin{user}
   build a model with five scalars: (3,1,-1/3), (3,1,-4/3), (3,2,7/6), (3,2,1/6), and (3,3,-1/3)
\end{user}

\begin{agent}
    I interpreted the BSM field content as scalar (3,1,-1/3), scalar (3,1,-4/3), scalar (3,2,7/6), scalar (3,2,1/6), scalar (3,3,-1/3). If this is correct, type only yes. If not, restate the quantum numbers.
\end{agent}

\begin{user}
   yes
\end{user}

\begin{agent}
    The model is constructed. [1.8s]
\end{agent}

\begin{user}
   pdf
\end{user}

\begin{agent}
\begin{verbatim}
    Generated report for 'BSM' with field content (3,1,-1/3), (3,1,-4/3),
    (3,2,7/6), (3,2,1/6), (3,3,-1/3). 
    TeX: /home/user/Downloads/bsm_agent/output/BSM.tex. 
    PDF: /home/user/Downloads/bsm_agent/output/BSM.pdf. 
    [1.2s]    
\end{verbatim}
\end{agent}

\begin{user}
   describe the model
\end{user}

\begin{agent}
\begin{verbatim}
Model: BSM     
Field content:
- phi_1 is a scalar in (3,1,-1/3): color triplet, weak singlet,
  hypercharge -1/3.
- phi_2 is a scalar in (3,1,-4/3): color triplet, weak singlet,
  hypercharge -4/3.
- phi_3 is a scalar in (3,2,7/6): color triplet, weak doublet,
  hypercharge 7/6.
- phi_4 is a scalar in (3,2,1/6): color triplet, weak doublet,
  hypercharge 1/6.
- phi_5 is a scalar in (3,3,-1/3): color triplet, weak triplet,
  hypercharge -1/3.

Interactions and consistency:
- Kinetic terms: 14
- Scalar potential: 73
- Yukawa: 14
- Mixed BSM interaction terms: 49
- Status: anomaly-free
- Anomaly summary: anomaly-free: su3_cubic=0, su2_su2_u1=0, su3_su3_u1=0, 
  u1_gravity=0, u1_cubic=0    
\end{verbatim}    
\end{agent}
This brief report once again confirms that the model is highly nontrivial, containing 14 Yukawa interaction terms, 73 scalar potential terms, and 14 gauge kinetic terms. Nevertheless, the package successfully generated the complete model in a matter of seconds.

\section{Conclusion}\label{sec:09}

We have introduced a framework that combines AI-driven natural-language interaction with symbolic BSM model building. The main novelty of this work is that the user only needs to specify the quantum numbers of the new fields in natural language, after which the framework automatically constructs and analyzes the corresponding renormalizable Standard Model extension. From this minimal input, the package performs the core symbolic stages of model building, including operator generation, anomaly checks, electroweak symmetry breaking, and tree-level mass-matrix extraction. The examples presented throughout the paper demonstrate that sophisticated model-building tasks can be carried out with substantially reduced user effort compared to traditional workflows.

More broadly, this work illustrates a practical paradigm for integrating large language models into scientific software. Rather than replacing established symbolic calculations, the LLM serves strictly as an interface and orchestration layer between the user and a deterministic computational engine. This separation preserves transparency, reproducibility, and scientific rigor while making advanced model-building tools more accessible to researchers and students. In this sense, \texttt{bsm_agent} fills a distinct niche in the BSM software ecosystem: it automates the pre-implementation stage of model building, transforming a specification of new field quantum numbers into a reproducible symbolic model analysis without requiring the user to manually enumerate the operator content of the theory. The package is freely available at
\href{https://github.com/saad-hep/bsm_agent}{\faGithub~bsm\_agent}
and is built entirely on open-source software.

\paragraph{Future directions.}
Future developments include support for even higher-dimensional SM representations beyond those currently implemented, additional gauge groups such as Pati--Salam, $SU(5)$, and $SO(10)$, incorporating discrete and continuous global symmetries, higher-dimensional operators, and the extension of the framework to supersymmetric theories. Further developments may also include more automated phenomenological analyses.

\subsection*{Acknowledgments}
SS acknowledges the financial support from the Slovenian Research Agency (research core funding No.~P1-0035 and project No.~N1-0321).

\appendix
\section{Quick Start} \label{summarization}

After installing the package (see Section~\ref{installation}), readers can immediately begin building new physics models. As an illustrative example, suppose a user wishes to construct a BSM model containing new fermions with quantum numbers $(1,3,1)+(1,3,-1)$ and a new scalar with quantum numbers $(1,4,3/2)$ under the SM gauge group. After launching the package (see Section~\ref{installation}),  the user can generate a complete model report using the following minimal set of chat interactions.

\begin{user}
fermion (1,3,1)+(1,3,-1) and scalar (1,4,3/2)    
\end{user}
\begin{agent}
   I interpreted the BSM field content as fermion (1,3,1), fermion (1,3,-1), scalar (1,4,3/2). If this is correct, type only yes. If not, restate the quantum numbers. 
\end{agent}
\begin{user}
yes
\end{user}
\begin{agent}
 The model is constructed. [0.1s]
\end{agent}
\begin{user}
generate pdf    
\end{user}
\begin{agent}
\begin{verbatim}
Generated report for 'SM+psi_1+psi_2+phi_1' with field content 
(1,3,1), (1,3,-1), (1,4,3/2). 
TeX: /home/user/Downloads/bsm_agent/output/SM_psi_1_psi_2_phi_1.tex. 
PDF: /home/user/Downloads/bsm_agent/output/SM_psi_1_psi_2_phi_1.pdf. 
[1.1s]
\end{verbatim} 
\end{agent}
\begin{user}
expand the lagrangian    
\end{user}
\begin{agent}
\begin{verbatim}
Generated report for 'SM+psi_1+psi_2+phi_1' with field content 
(1,3,1), (1,3,-1), (1,4,3/2). 
TeX: /home/user/Downloads/bsm_agent/output/SM_psi_1_psi_2_phi_1.tex. 
PDF: /home/user/Downloads/bsm_agent/output/SM_psi_1_psi_2_phi_1.pdf. 
[2.6s]    
\end{verbatim}
\end{agent}
\begin{user}
generate mass matrix
\end{user}
\begin{agent}
\begin{verbatim}
Generated report for 'SM+psi_1+psi_2+phi_1' with field content 
(1,3,1), (1,3,-1), (1,4,3/2). 
TeX: /home/user/Downloads/bsm_agent/output/SM_psi_1_psi_2_phi_1.tex. 
PDF: /home/user/Downloads/bsm_agent/output/SM_psi_1_psi_2_phi_1.pdf. 
[4.7s]    
\end{verbatim} 
\end{agent}
\begin{user}
summarize the model 
\end{user}
\begin{agent}
\begin{verbatim}
Model: SM+psi_1+psi_2+phi_1
Field content:
- psi_1 is a fermion in (1,3,1): color singlet, weak triplet, hypercharge 1.
- psi_2 is a fermion in (1,3,-1): color singlet, weak triplet, hypercharge -1.
- phi_1 is a scalar in (1,4,3/2): color singlet, weak quartet, hypercharge 3/2.

Interactions and consistency:
  - Kinetic terms: 12
  - Fermion mass: 1
  - Scalar potential: 8
  - Yukawa: 5
  - Mixed BSM interaction terms: 2
  - Status: anomaly-free
  - Anomaly summary: anomaly-free: su3_cubic=0, su2_su2_u1=0, su3_su3_u1=0,
  u1_gravity=0, u1_cubic=0    
\end{verbatim} 
\end{agent}

\bibliographystyle{style}
\bibliography{reference}
\end{document}